\documentclass[twocolumn, usenatbib]{aastex61}

\submitjournal{ApJ}

\shortauthors{Ryon et al.}
\begin{document}

\title{Effective Radii of Young, Massive Star Clusters in Two LEGUS Galaxies\footnote{Based on observations obtained with the NASA/ESA Hubble Space Telescope, at the Space Telescope Science Institute, which is operated by the Association of Universities for Research in Astronomy, Inc., under NASA contract NAS 5-26555. These observations are associated with program \#13364.}}

\correspondingauthor{Jenna E. Ryon}
\email{ryon@stsci.edu}

\author{J. E. Ryon}
\affiliation{University of Wisconsin-Madison, 475 N. Charter St., Madison, WI 53706}
\affiliation{Space Telescope Science Institute, 3700 San Martin Dr., Baltimore, MD 21218}

\author{J. S. Gallagher}
\affiliation{University of Wisconsin-Madison, 475 N. Charter St., Madison, WI 53706}

\author{L. J. Smith}
\affiliation{European Space Agency/Space Telescope Science Institute, 3700 San Martin Dr., Baltimore, MD 21218}

\author{A. Adamo} 
\affiliation{Dept. of Astronomy, The Oskar Klein Centre, Stockholm University, Stockholm, Sweden}

\author{D.Calzetti}
\affiliation{Dept. of Astronomy, University of Massachusetts -- Amherst, Amherst, MA 01003}

\author{S. N. Bright}
\affiliation{Space Telescope Science Institute, 3700 San Martin Dr., Baltimore, MD 21218}

\author{M. Cignoni}
\affiliation{Space Telescope Science Institute, 3700 San Martin Dr., Baltimore, MD 21218}
\affiliation{Department of Physics, University of Pisa, Largo Pontecorvo, 3 Pisa, I-56127, Italy}

\author{D. O. Cook}
\affiliation{Dept. of Physics and Astronomy, University of Wyoming, Laramie, WY}
\affiliation{California Institute of Technology, 1200 East California Blvd, Pasadena, CA 91125}

\author{D. A. Dale}
\affiliation{Dept. of Physics and Astronomy, University of Wyoming, Laramie, WY}

\author{B. E. Elmegreen}
\affiliation{IBM Research Division, T.J. Watson Research Center, Yorktown Hts., NY}

\author{M. Fumagalli}
\affiliation{Institute for Computational Cosmology and Centre for Extragalactic Astronomy, Durham University, Durham, United Kingdom}

\author{D. A. Gouliermis}
\affiliation{Zentrum f\"ur Astronomie der Universit\"at Heidelberg, Institut f\"ur Theoretische Astrophysik, Albert-Ueberle-Str.\,2, 69120 Heidelberg, Germany}
\affiliation{Max Planck Institute for Astronomy,  K\"{o}nigstuhl\,17, 69117 Heidelberg, Germany}

\author{K. Grasha}
\affiliation{Dept. of Astronomy, University of Massachusetts -- Amherst, Amherst, MA 01003}

\author{E. K. Grebel}
\affiliation{Astronomisches Rechen-Institut, Zentrum f\"ur Astronomie der Universit\"at Heidelberg, M\"onchhofstr.\ 12--14, 69120 Heidelberg, Germany}

\author{H. Kim}
\affiliation{Gemini Observatory, Casilla 603, La Serena, Chile}

\author{M. Messa}
\affiliation{Dept. of Astronomy, The Oskar Klein Centre, Stockholm University, Stockholm, Sweden}

\author{D. Thilker}
\affiliation{Dept. of Physics and Astronomy, The Johns Hopkins University, Baltimore, MD}

\author{L. Ubeda}
\affiliation{Space Telescope Science Institute, 3700 San Martin Dr., Baltimore, MD 21218}

\begin{abstract}
We present a study of the effective (half-light) radii and other structural properties of a systematically selected sample of young, massive star clusters (YMCs, $\geq$$5\times10^3$~M$_{\odot}$ and $\leq$200~Myr) in two nearby spiral galaxies, NGC~628 and NGC~1313. We use Hubble Space Telescope WFC3/UVIS and archival ACS/WFC data obtained by the Legacy Extragalactic UV Survey (LEGUS), an HST Treasury Program. We measure effective radii with GALFIT, a two-dimensional image-fitting package, and with a new technique to estimate effective radii from the concentration index (CI) of observed clusters. The distribution of effective radii from both techniques spans $\sim$0.5--10~pc and peaks at 2-3~pc for both galaxies. We find slight positive correlations between effective radius and cluster age in both galaxies, but no significant relationship between effective radius and galactocentric distance. Clusters in NGC~1313 display a mild increase in effective radius with cluster mass, but the trend disappears when the sample is divided into age bins. We show that the vast majority of the clusters in both galaxies are much older than their dynamical times, suggesting they are gravitationally bound objects. We find that about half of the clusters in NGC~628 are underfilling their Roche lobes, based on their Jacobi radii. Our results suggest that the young, massive clusters in NGC~628 and NGC~1313 are expanding due to stellar mass loss or two-body relaxation and are not significantly influenced by the tidal fields of their host galaxies.
\end{abstract}


\section{Introduction}
\label{intro}

The discovery of young, massive clusters (YMCs) residing in nearby galaxies has spurred major interest in recent decades in determining their properties and evolution. The sizes of YMCs appear to be nearly constant across a wide range of age, mass, and environment \citep[e.g.,][]{portegieszwart2010}. The radius containing half of the total cluster light, the effective radius or $r_{\mathrm{eff}}$, is the most straightforward size scale to measure observationally, and is typically found to be 2-3~pc for YMCs \citep[e.g.,][]{elson1987,whitmore1999,larsen2004, barmby2006,scheepmaker2007,portegieszwart2010, bastian2012a}. Interestingly, globular clusters (GCs) also have characteristic effective radii of 2-3~pc \citep[e.g.,][]{jordan2005,harris2009, masters2010, puzia2014}. If today's YMCs are modern-day progenitors of ancient GCs, then studying YMC sizes and evolution locally may shed light on the origins of GCs.

The size of a star cluster is tied to the internal and external mechanisms which influence the dynamical state of the stars in the cluster. A better understanding of the sizes of YMCs can therefore constrain their formation and early evolution. For instance, the relationship between cluster age and radius may show whether YMCs of a certain age and mass range are expanding or contracting, which would indicate whether they are behaving as isolated or tidally-limited systems  \citep{heggiehut2003,Trenti10,alexander2013}. Similarly, the dependence of cluster radius on distance from the galaxy center may provide clues to the influence of the galaxy's tidal field on YMC evolution  \citep{gieles2011b,madrid2012,alexander2013,Sun16}. The relationship between cluster mass and radius may illuminate the effect of perturbations by giant molecular clouds (GMCs). For a weak mass-radius relationship, which has been found by several studies  \citep[e.g.,][]{zepf1999, larsen2004, scheepmaker2007,barmby2009}, less massive objects are more likely to be disrupted by GMC interactions because they are of lower density \citep{gieles2006c}. In addition, \cite{gieles2011a} show it is possible to determine whether an object is likely to be gravitationally bound (star clusters) or unbound (associations) by comparing the crossing time, calculated from mass and radius, to the object's age.

One of the central goals of the Legacy Extragalactic UV Survey \citep[LEGUS;][]{calzetti2015} is to better understand the role of star clusters in the star formation process. Part of this goal is to determine the shape of the cluster radius distribution and whether it depends on galaxy environment. In this paper, we directly address this goal by studying the sizes of homogeneously-selected YMCs in two LEGUS galaxies, NGC~628 and NGC~1313. These two galaxies were chosen for this study because of their relatively numerous cluster populations and differing morphological types. They provide an interesting contrast to probe effects of the environment on cluster structure. In addition, NGC~1313 is half the distance of NGC~628, and we can therefore test  how spatial resolution affects the measured cluster properties.

NGC~628 (M74) is a face-on ($i=25.2^{\circ}$) grand-design spiral galaxy (SAc) located at a distance of 9.9$\pm$1.3~Mpc \citep{olivares2010}. It has a stellar mass of $1.1\times10^{10}$~M$_{\odot}$ and an extinction-corrected UV star formation rate (SFR) of 3.67~M$_{\odot}$/yr \citep{calzetti2015}. \cite{thilker2007} noted the presence of an extended UV disk featuring a spiral structure that is a continuation of the inner, optically-bright pattern. Though it is the largest member of a galaxy group, the regular appearance of its disk suggests no recent interactions. \cite{adamo2017} provide an overview of the cluster analysis techniques employed by LEGUS and presents results on the luminosity function, mass function, and age distribution of the YMC population of NGC~628 as a test case. \cite{larsen2004} measured effective radii for 30 clusters in NGC~628, and found an average radius of 3.65$\pm$0.55~pc.

 NGC~1313 is a somewhat inclined ($i=40.7^{\circ}$) SBd galaxy located at a distance of 4.39$\pm$0.04~Mpc \citep{jacobs2009}. Its stellar mass is $2.6\times10^9$~M$_{\odot}$ and the extinction-corrected UV SFR is 1.15~M$_{\odot}$/yr \citep{calzetti2015}. The resemblance between NGC~1313 and the Large Magellanic Cloud has been noted previously \citep{devaucouleurs1963}, given its bar and rather irregular appearance. A number of studies suggest that NGC~1313 may be interacting with a satellite galaxy which has produced a loop of H{\,\sc i} gas around the galaxy \citep{peters1994} and led to an increase in star formation rate in the southwestern part of the galaxy over the past 100~Myr \citep{silvavilla2012}. Previous studies have noted strong evidence for disruption of young clusters \citep{pellerin2007}, a high cluster formation rate \citep{silvavilla2011}, and cluster radii between $\sim$2 and 5~pc, on average \citep{larsen2004,mora2009}.

This work builds upon the techniques and results from  \cite[][hereafter Paper~I]{ryon2015}, in which we measured the effective radii and light profile slopes of $\sim$200 YMCs in the nearby spiral galaxy M83 using GALFIT, a two-dimensional image fitting package. In this paper, we select clusters from  two adjacent \textit{HST} fields obtained by LEGUS for each of NGC~628 and NGC~1313, resulting in samples of 320 and 195 YMCs, respectively. We fit these clusters with GALFIT to determine their effective radii and light profile slopes.  Since GALFIT does not properly fit some types of clusters, we also calculate an estimate of the effective radius from the concentration index (CI) of each cluster using a relation determined from artificial clusters. We compare the effective radii and light-profile shapes of the clusters to their ages, masses, and galactocentric distances to probe the mechanisms that drive their structural evolution, and further investigate their dynamical states.

This paper is organized as follows. In Section~\ref{obs-cat}, we describe the observations and star cluster catalog. We describe our methods for measuring effective radii and completeness tests in Section~\ref{methods}. In Section~\ref{results}, we present the results of our measurements and explore relationships between size and other cluster properties. We briefly discuss the implications of this work and summarize our findings in Section~\ref{discussion-conclusions}.

\section{Observations and  Cluster Catalog}
\label{obs-cat}

LEGUS is a \textit{Hubble Space Telescope} (HST) Cycle 21 Treasury program, which obtained imaging of 50 nearby galaxies (within $\sim$13 Mpc) in five filters with WFC3/UVIS and ACS/WFC. New imaging for selected pointings on each galaxy were obtained with WFC3/UVIS to complement archival ACS/WFC imaging and complete the multiband coverage from the near-UV to the I-band. All images are drizzled to the UVIS native pixel scale of 0$''$.03962/pixel. See \cite{calzetti2015} for a complete description of the data reduction of the LEGUS imaging datasets. In this study, we measure star cluster sizes from the F555W images\footnote{The final reduced images are available at \doi{10.17909/T9J01Z}.}.

The production of catalogs of candidate star clusters by LEGUS is described in detail in \cite{adamo2017}. Here, we briefly describe the catalogs from which the cluster samples for this study were selected.

Separate catalogs are produced for each pointing on the two galaxies: NGC~628c (central pointing), NGC~628e (east pointing), NGC~1313e (east pointing), and NGC~1313w (west pointing). First, SourceExtractor \citep{bertin1996} identifies sources in the white-light image of each pointing   \cite[see][for a description of the white-light images for LEGUS]{calzetti2015}. Next, growth curves and CI values are determined for user-identified isolated stars and star cluster candidates, which  allow selection of the appropriate photometric aperture size and CI value for separating stars and cluster candidates. The CI is the magnitude difference between aperture radii of 1 and 3 pixels in the  F555W-band image, and is therefore larger for more extended objects \citep[e.g.,][]{holtzman1996,whitmore2010}. Aperture photometry is performed in all five filters using the science aperture radius determined from the isolated clusters and a  background annulus located at 7~pixels with a width of 1~pixel. 

 Average aperture corrections are calculated from the isolated clusters by measuring the average magnitude difference from the science aperture to a radius of 20 pixels in each band. We apply the aperture corrections and Galactic foreground extinction corrections to the science photometry. Finally, for each pointing, a catalog for visual inspection is produced by performing a series of cuts: each source must have a CI value larger than the CI limit determined from isolated stars and clusters, be detected in at least four bands with a photometric error below 0.3~mag, and have an absolute  F555W-band magnitude brighter than $-6$~mag. Relevant parameters used by the LEGUS team to produce the candidate cluster catalogs for each pointing are provided in Table~\ref{catalog}.

\begin{table}
\centering
\caption{LEGUS Cluster Catalog Parameters}
\label{catalog}
\begin{tabular}{ccccc}
\hline
&Camera & CI &  Aperture & Distance \\
&F555W & Limit & Radius & \\
&& & (pix) & (Mpc)\\
\hline
NGC~628c & ACS & 1.4 & 4 & 9.9\\
NGC~628e & WFC3 & 1.3 & 4 & 9.9 \\
NGC~1313e & ACS & 1.4 & 6 & 4.39 \\
NGC~1313w & ACS & 1.4 & 6 & 4.39\\
\hline
\end{tabular}
\end{table}

At least three members of the LEGUS team visually inspect each cluster candidate that satisfies the above criteria. The cluster candidate is assigned one of four classes by each LEGUS team member. The descriptions of each class are as follows:

\begin{description}

\item [Class 1] Compact and centrally concentrated with a FWHM more extended than that of a star. Homogeneous in color.

\item [Class 2] Slightly elongated or asymmetric light profile shapes with a FWHM more extended than that of a star. Homogeneous in color.

\item [Class 3] Asymmetric light profiles consisting of multiple peaks on top of diffuse underlying light.

\item [Class 4] Spurious sources including single stars, pairs of stars (color difference), chip edge artifacts, hot pixels, and background galaxies.

\end{description}
The mode and mean class is determined for each source and listed in the final LEGUS catalog. 

 The age, mass, and extinction of each cluster candidate in the visual inspection catalogs with photometric detections in at least four filters are determined using the SED-fitting code \textit{Yggdrasil} \citep{zackrisson2011}. For this study, we use the catalogs containing fits performed with Padova-AGB isochrones, available in Starburst99 \citep{leitherer1999,vazquez2005}, and the Milky Way extinction law from \cite{cardelli1989}. The results of the SED-fitting are listed in the final cluster candidate catalogs together with ID numbers, RA Dec and pixel coordinates, magnitudes, CI values, residuals from the SED-fitting, reduced chi-squared of the SED fits, and visual inspection class assignments.

\begin{figure}
\centering
\includegraphics[width=\columnwidth]{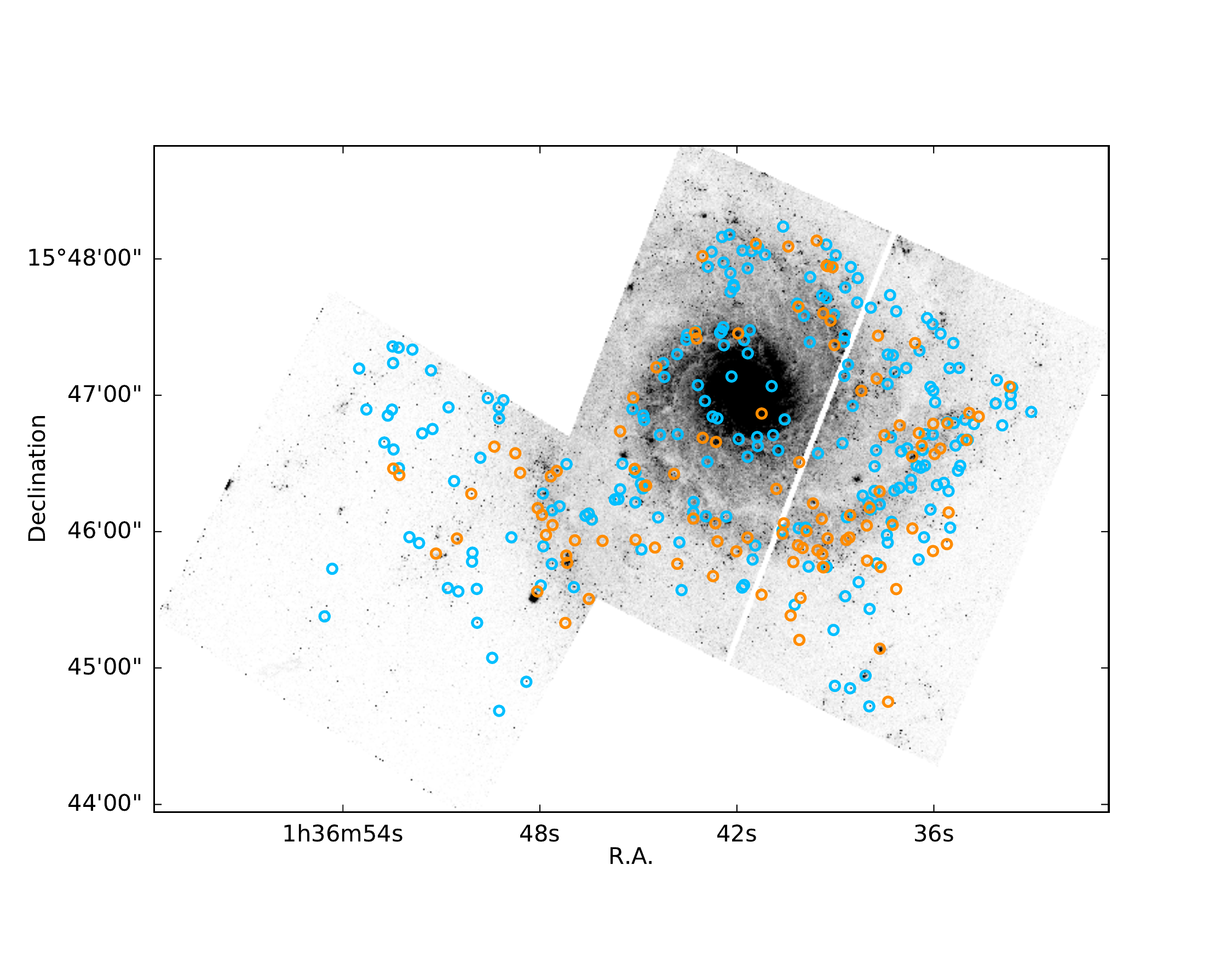}
\caption{Grayscale F555W image of NGC~628 with the locations of cluster candidates overplotted as open circles. Class 1 objects are blue circles and class 2 objects are orange circles. \label{cluster_loc_628}}
\end{figure}

\begin{figure}
\centering
\includegraphics[width=\columnwidth]{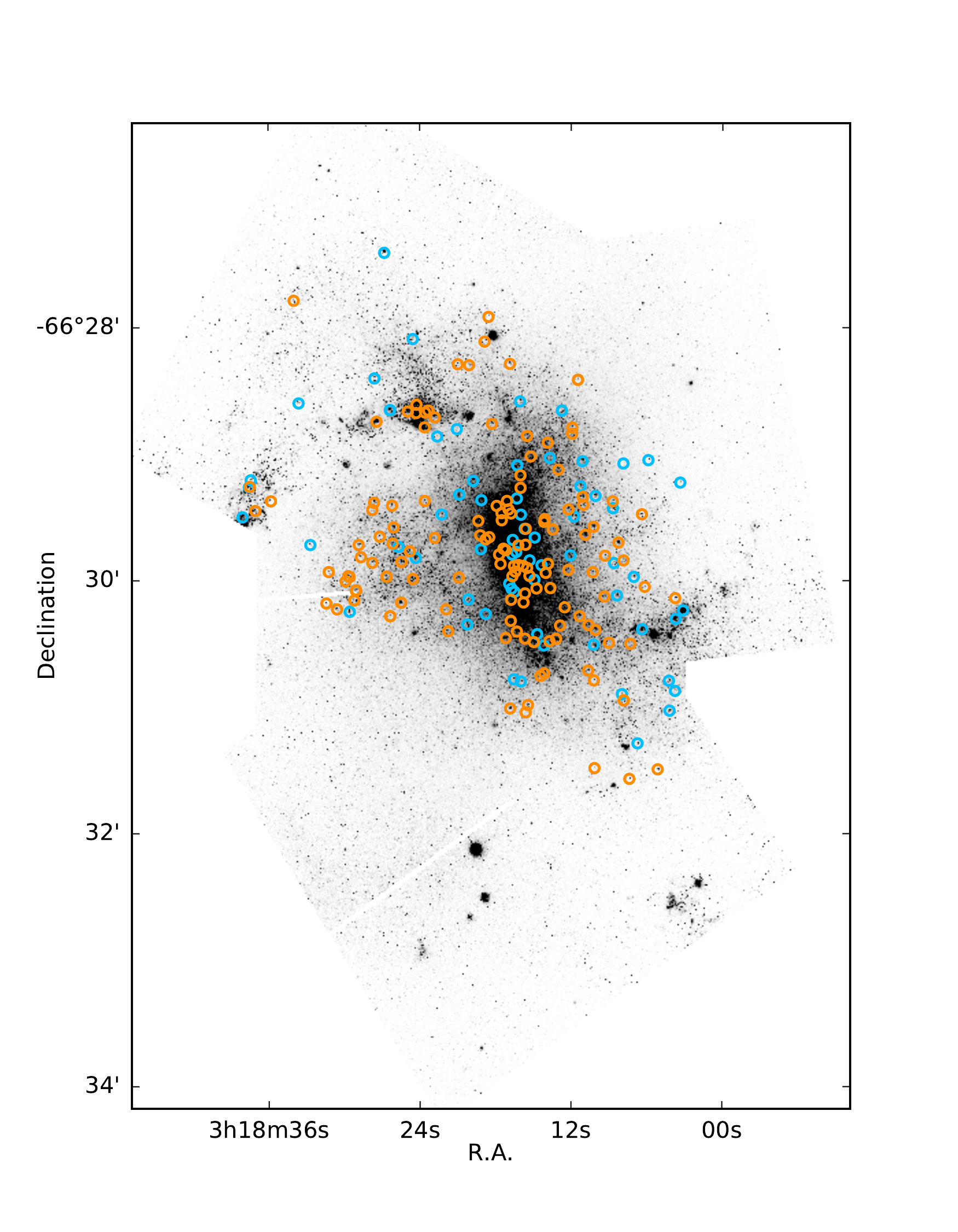}
\caption{Same as for Figure~\ref{cluster_loc_628}, but for NGC~1313. \label{cluster_loc_1313}}
\end{figure}

 We impose further selection criteria to select a bona fide sample of young, massive star cluster candidates from both galaxies. First, we limit our samples to objects with a mode visual inspection class of 1 or 2. This ensures that the light profiles are relatively well-behaved, and that the objects are centrally-concentrated, as  is expected for gravitationally-bound star clusters. We also limit our sample to objects with masses $\geq$5000~M$_{\odot}$ and ages $\leq$200~Myr.  The mass cut minimizes the effects of stochastic sampling of the stellar IMF \citep[e.g.,][]{popescu2010,fouesneau2010}, and the age cut ensures that we are not strongly affected by incomplete detection of clusters due to evolutionary fading. In Figures~\ref{cluster_loc_628} and \ref{cluster_loc_1313}, we plot the locations of clusters that satisfy these selection criteria as open circles on grayscale F555W mosaic images of NGC~628 and NGC~1313, respectively. The blue circles are visual inspection class 1 cluster candidates, and the orange circles are class 2.

\begin{figure*}
\centering
\includegraphics[width=0.7\textwidth]{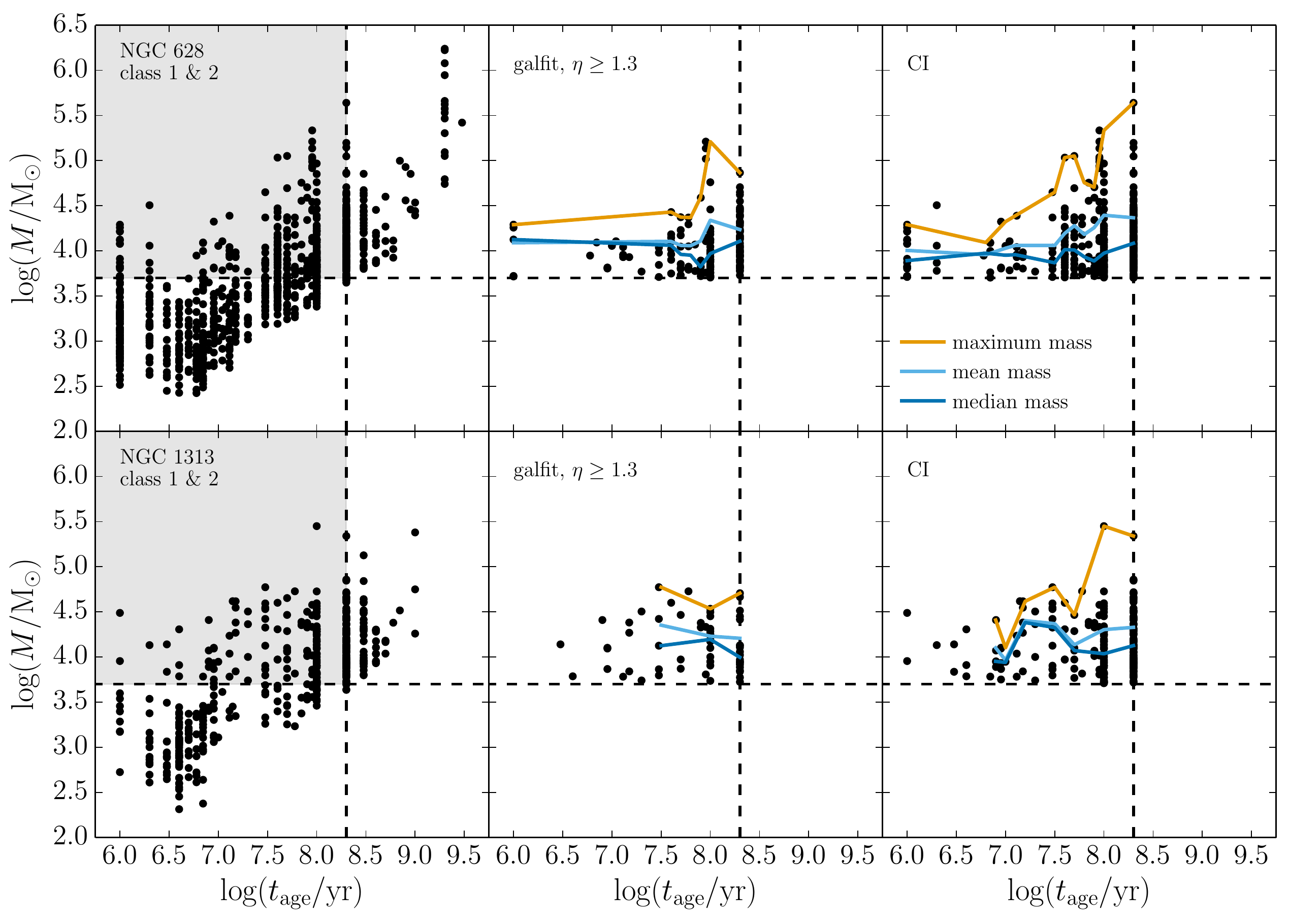}
\caption{Cluster candidate ages and masses for NGC~628 (top row) and NGC~1313 (bottom row), as determined from SED fits. \textit{Left:} All class 1 and 2 objects. The dashed lines represent locations of age and mass cuts, and the shaded region contains the cluster sample for which we measure sizes. Clusters $\leq$200~Myr and $\geq$5$\times10^3$~M$_{\odot}$ are included. \textit{Center:} Class 1 and 2 objects satisfying the age and mass cuts that are also successfully fit by GALFIT and are best described by a power-law light profile slope of $\eta \geq 1.3$. As discussed in Section~\ref{reff}, slopes shallower than $\eta = 1.3$ make it difficult to accurately constrain $r_{\mathrm{eff}}$. The yellow, light blue, and dark blue lines represent the maximum, mean, and median mass, respectively, in bins of width 0.1 in log-space that contain 5 or more objects. \textit{Right:} Class 1 and 2 objects satisfying the age and mass cuts for which effective radii have been estimated from CI values. Lines are the same as in the center panel. \label{age-mass}}
\end{figure*}

In Figure~\ref{age-mass}, we plot the ages and masses of cluster candidates in NGC~628 (top row) and NGC~1313 (bottom row) as determined from the SED fits. The left panels show the class 1 and 2 objects in each galaxy. The dashed lines represent the age and mass cuts we have applied to the samples, and the shaded regions contain all objects that we have attempted to fit with GALFIT and estimated effective radii from the CI. The middle panels show the clusters successfully fit with GALFIT, and the right panels show the objects for which effective radii have been estimated from CI values. The yellow, light blue, and dark blue solid lines show the maximum, mean, and median cluster mass, respectively, in bins of width 0.1 dex in age that contain at least 5 objects.   These data show the expected statistical correlation between cluster age and mass that results from the larger number of older star clusters which more completely sample the cluster upper mass range.

The total number of clusters remaining in the NGC~628 sample after the visual inspection class, mass, and age cuts have been applied is 320, which consists of 257 in the central pointing and 63 in the east pointing. For NGC~1313, the total number of clusters remaining is 195, which consists of 54 in the east pointing and 141 in the west pointing.

\section{Methods}
\label{methods}

\subsection{Effective Radii from GALFIT}
\label{galfit}

The methods used for fitting the two-dimensional light profiles of cluster candidates with GALFIT \citep{peng2002, peng2010} are described in detail in Paper~I. We briefly summarize those methods and a few small differences here.

Other studies have used ISHAPE \citep{larsen1999b} to measure the structural properties of star clusters in nearby galaxies. A comparison of the results from ISHAPE and GALFIT fits of a subsample of well-behaved clusters included in Paper~I found differences in measured effective radii on the order of about 6\%. It is therefore unlikely that the choice of fitting software would strongly affect the results we present here. We chose to use GALFIT because its reporting of fitting results and errors is somewhat clearer than ISHAPE.

To prepare the HST F555W images for use with GALFIT, we multiplied the drizzled images by the exposure time to convert from units of $\mathrm{e^-}$/s to $\mathrm{e^-}$. We also updated the image headers to set the GAIN keyword equal to 1.0~$\mathrm{e^-}$/ADU and to include the readnoise for the appropriate camera, 3.11~$\mathrm{e^-}$ for WFC3/UVIS and 4.2~$\mathrm{e^-}$ for ACS/WFC, in the header keyword RDNOISE.

In order to extract the structural components of each cluster from the LEGUS images, GALFIT convolves a model image with a point spread function (PSF) and compares the result to the observed data. An accurate stellar PSF is essential for reproducing the effects of the telescope optics in the model images. We create PSFs for each LEGUS pointing from several bright, isolated stars in each image by using \texttt{pstselect} and \texttt{psf} within DAOPHOT in IRAF. We spatially subsample the empirical PSFs by a factor of 10.

We assume an EFF light profile shape (also known as a Moffat profile), because it describes well the light profiles of young star clusters in particular \citep{elson1987, larsen1999b, mclaughlin2005}. The EFF profile takes the form
\begin{equation}
\mu(r) = \mu_{0}(1+r^2/a^2)^{-\eta}
\end{equation}
where $\mu$ is the surface brightness, $a$ is a characteristic radius, and $\eta$ is the power-law exponent of the profile wings. Note that $\eta$ is equal to $\gamma/2$ in Equation~1 of \cite{elson1987}. The effective radius ($r_{\mathrm{eff}}$), or half-light radius, is defined to be the radius of the circle which contains half of the total surface brightness of the light profile. This is written as
\begin{equation} \label{eqreff}
 r_{\mathrm{eff}} = \mathrm{FWHM}\ \frac{\sqrt{(1/2)^{\frac{1}{1-\eta}} - 1}}{2\sqrt{2^{1/\eta} -1}}, 
\end{equation}
which is only valid for $\eta>1.0$. For an elliptical profile, the true effective radius can be found by multiplying Equation~\ref{eqreff} by a factor of $0.5(1+b/a)$, where $b/a$ is the semiminor to semimajor axis ratio  supplied by GALFIT (ISHAPE manual, \citealt{larsen1999b}).

For each cluster, both an EFF light profile and a local background component are fit simultaneously over a $30 \times 30$~pixel region centered on the cluster. This fitting region size was selected to be consistent with that used in Paper~I for M83. Different fitting region sizes were tested in Paper~I and found to not significantly affect the overall results (see Table~2 in that work). The fitting region is equivalent to $\approx57 \times 57$~pc at the distance of NGC~628 and $\approx25 \times 25$~pc at the distance of NGC~1313. The amplitude of the local background is left as a free parameter. The free parameters for the EFF component are the x and y image coordinates of the cluster center, total magnitude, FWHM, $\eta$, axis ratio ($b/a$), and position angle. Table~\ref{input} lists the initial guesses for each of the free parameters, excluding the xy coordinates, for both galaxies. GALFIT returns the best-fit values for each free parameter and their $1\sigma$ uncertainties. At times, GALFIT cannot converge on a best-fitting model, or one of the fitting parameters becomes unphysically small. The objects for which this occurred, of which there were 46 in the NGC~628 sample and 29 in the NGC~1313 sample, are labeled in the GALFIT output file. We exclude them from the GALFIT sample.

\begin{table}
\centering
\caption{GALFIT Input Parameters}
\label{input}
\begin{tabular}{cc}
\hline
Parameter & Value \\
\hline
Total Magnitude & 20.0 mag,17.0 mag\tablenotemark{a} \\
FWHM & 2.5 pix \\
$\eta$ & 1.5 \\
Axis Ratio & 1.0 \\
Position Angle & $25^{\circ}$ \\
Background & 300.0 $\mathrm{e^{-}}$ \\
\hline
\end{tabular}
\tablenotetext{a}{The total magnitude initial guess is 20.0~mag for NGC~628 and 17.0~mag for NGC~1313.}
\end{table}

Finally, we inspect the residuals image produced by GALFIT for each cluster candidate. Clusters for which the EFF component of the fit appears to be influenced by the presence of other objects within the fitting region (e.g., stars, clusters, areas of high background) are flagged and excluded from the following analysis. Of the clusters in the NGC~628 sample, 58 were found to be affected by other objects within the fitting region. In NGC~1313, 64 clusters were affected. For some of the objects, GALFIT also could not converge on a best-fitting solution, as described above. Combining the selection criteria listed in this section and Section~\ref{obs-cat}, the total number of clusters successfully fit with GALFIT can be determined. There are 241 clusters remaining in the NGC~628 sample, including 194 in the central pointing and 47 in the east pointing, and 130 remaining in the NGC~1313 sample, including 36 in the east pointing and 94 in west pointing.

\subsection{Effective Radii Estimated from Concentration Index}
\label{ci-estimated}

In addition to measuring effective radii using GALFIT, we have developed a method to estimate effective radii from a cluster's CI. To do this, we create artificial clusters with the LEGUS cluster completeness tool described in \cite{adamo2017}. The first step of this tool uses routines in the \texttt{baolab} environment \citep{larsen1999b} to create artificial star clusters from an input stellar PSF. The PSF is convolved with  a symmetric EFF light profile with a power-law index of $\eta = 1.5$ and a pre-determined effective radius to produce an artificial cluster of a given size. For each input radius, a single frame is created, containing 500 artificial clusters of that radius and a range of apparent magnitudes. The magnitude ranges, which are $\sim$18 to 24~mag for NGC~628 and $\sim$17 to 23~mag for NGC~1313 in F555W, were chosen to match the magnitude range of the real clusters in each galaxy. The artificial clusters are placed randomly within the frame, which is then added to a defined region within the F555W image of a LEGUS pointing. This region is determined by finding where the WFC3/UVIS and ACS/WFC footprints overlap for each pointing. We produce images containing artificial clusters with effective radii of 0.5~pc to 15.0~pc in steps of 0.5~pc, as well as one frame of 0.25~pc radius clusters for each of the NGC~1313 pointings.

\begin{figure}
\centering
\includegraphics[width=\columnwidth]{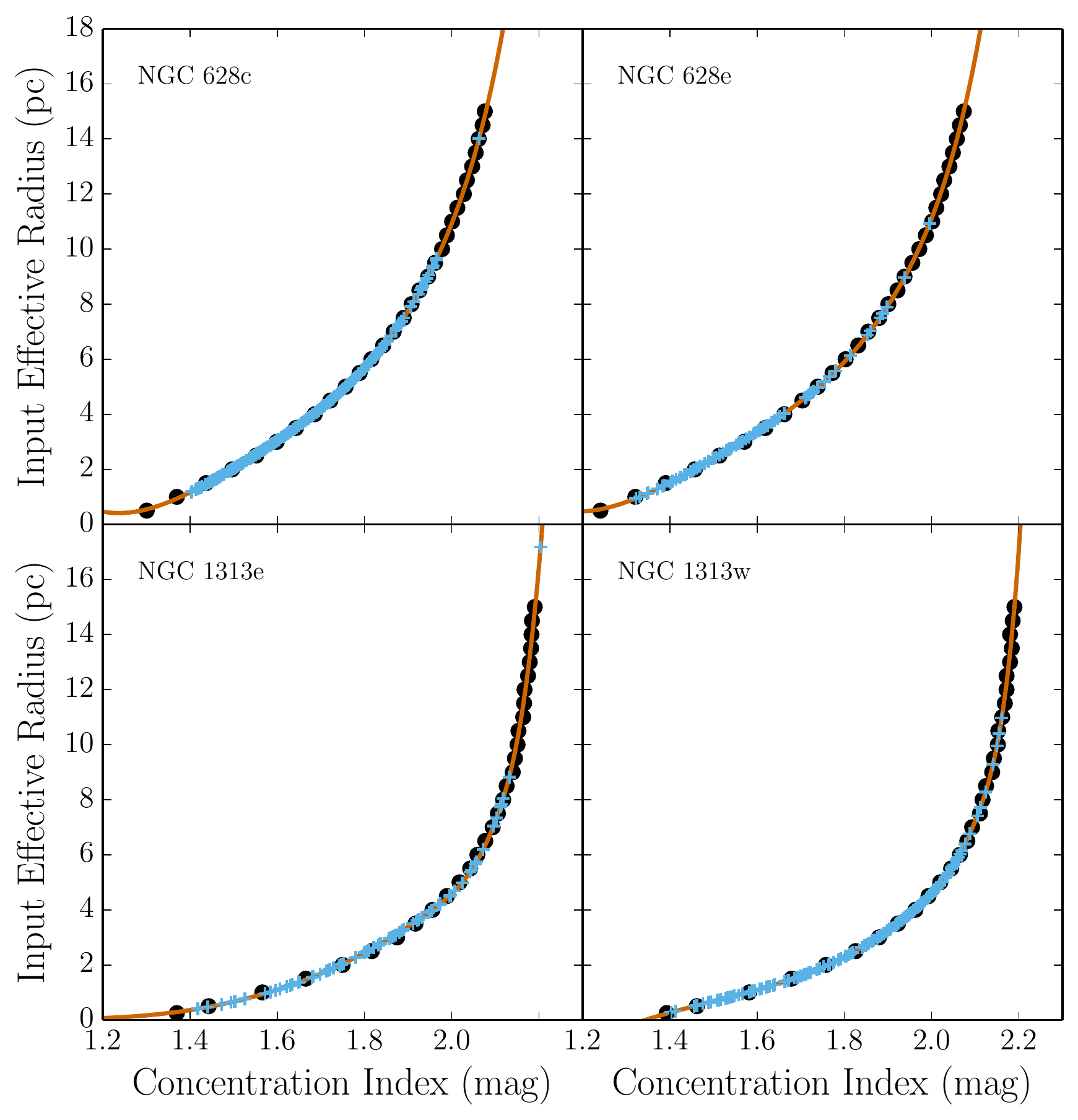}
\caption{Input effective radius versus measured CI for artificial clusters in the four LEGUS pointings. The black points show the median CI values of 500 artificial clusters with the same input effective radius. The orange line is the univariate spline fit to the black points. The blue crosses show the estimated effective radii of observed clusters, calculated from the spline fit. \label{ci-reff-rel} }
\end{figure}

For each LEGUS pointing, we measure the CI of each artificial cluster in every frame using the same procedure as for the observed clusters. We calculate the median and median absolute deviation \citep[MAD,][]{feigelson2012} of the CI  of the artificial clusters with each input effective radius. The range of median CI values of the artificial clusters matches that of the observed clusters well. For the NGC~628 pointings, we use the \texttt{UnivariateSpline} class within \texttt{scipy.interpolate} to fit a fourth-order univariate spline with a smoothing factor of $s=0.1$ to the median CI and input radius values. For the NGC~1313 pointings, we use the same python class to fit a third-order univariate spline with a smoothing factor of $s=1.2$ to the median CI and input radius values. In Figure~\ref{ci-reff-rel}, we plot the median CI and input effective radii of the artificial clusters along with the spline fit for each of the four pointings. 

From this relation, we are able to estimate effective radii from the CI values of the observed clusters in each pointing.  The small level of scatter in Figure~\ref{ci-reff-rel} indicates that the CI fits are not overly sensitive to cluster ellipticity. This is consistent with other studies that show circular profiles give good results especially for cases where the axial ratio $b/a$ is greater than $\sim$0.3 \citep[e.g.,][]{Matthews99,Smith01}. The difference in the overall shapes of the CI-$r_{\mathrm{eff}}$ relationships between the two galaxies is likely a distance effect. 

 Artificial clusters of the same input radii have larger CI values in NGC~1313 than NGC~628 because the former galaxy is approximately half the distance of latter. We estimate 1$\sigma$ errors on the effective radii by creating 5000 Monte Carlo realizations of the observed CI values assuming the CI photometric errors are 1$\sigma$ uncertainties and calculating the standard deviation of the resulting effective radius distribution for each observed cluster. Using this method, we are able to estimate effective radii for all of the clusters that satisfy the selection criteria listed in Section~\ref{obs-cat}, i.e., 320 clusters in NGC~628 and 195 in NGC~1313. As shown by the blue crosses in Figure~\ref{ci-reff-rel}, only a small fraction of observed clusters have CI values greater than 2.0. Above this value, the CI-effective radius relations become quite steep, and therefore a small uncertainty in CI leads to a large uncertainty in effective radius.

\subsection{Completeness Tests}
\label{completeness}

\subsubsection{Artificial Cluster Tests}
\label{artificial}

We perform tests on the artificial clusters created in Section~\ref{ci-estimated} in order to determine if extremely compact star clusters are preferentially lost from the sample due to problems with convergence of GALFIT fits. To do this, we run GALFIT on artificial clusters in the F555W images of each pointing on NGC~628 and NGC~1313, which were created as described in Section~\ref{ci-estimated}. We consider a subset of those images containing artificial clusters with effective radii ranging between 0.5 and 15~pc for NGC~628 and 0.25 and 15~pc for NGC~1313. The initial guesses for the GALFIT fits are the same as for the observed star clusters (see Table~\ref{input}).

Next, we perform aperture photometry on the artificial clusters in the same manner as for the observed star clusters. For bins in magnitude of width 0.5~mag, we find the number of artificial clusters that are ``recovered'' by GALFIT, which means that their fits successfully converged. This is done for each input effective radius. We then divide the number of recovered clusters by the total to get the percentage of objects recovered in each magnitude bin for every effective radius. Photometric blends were not removed.
We calculate errors by taking the square root of the number of clusters ``recovered'' in each bin, dividing by the total number input into GALFIT, and converting to a percentage. The total number of clusters input into GALFIT summing over all magnitude bins is 500 and the errors become very large when small numbers of clusters are recovered.

\begin{figure*}
\gridline{\fig{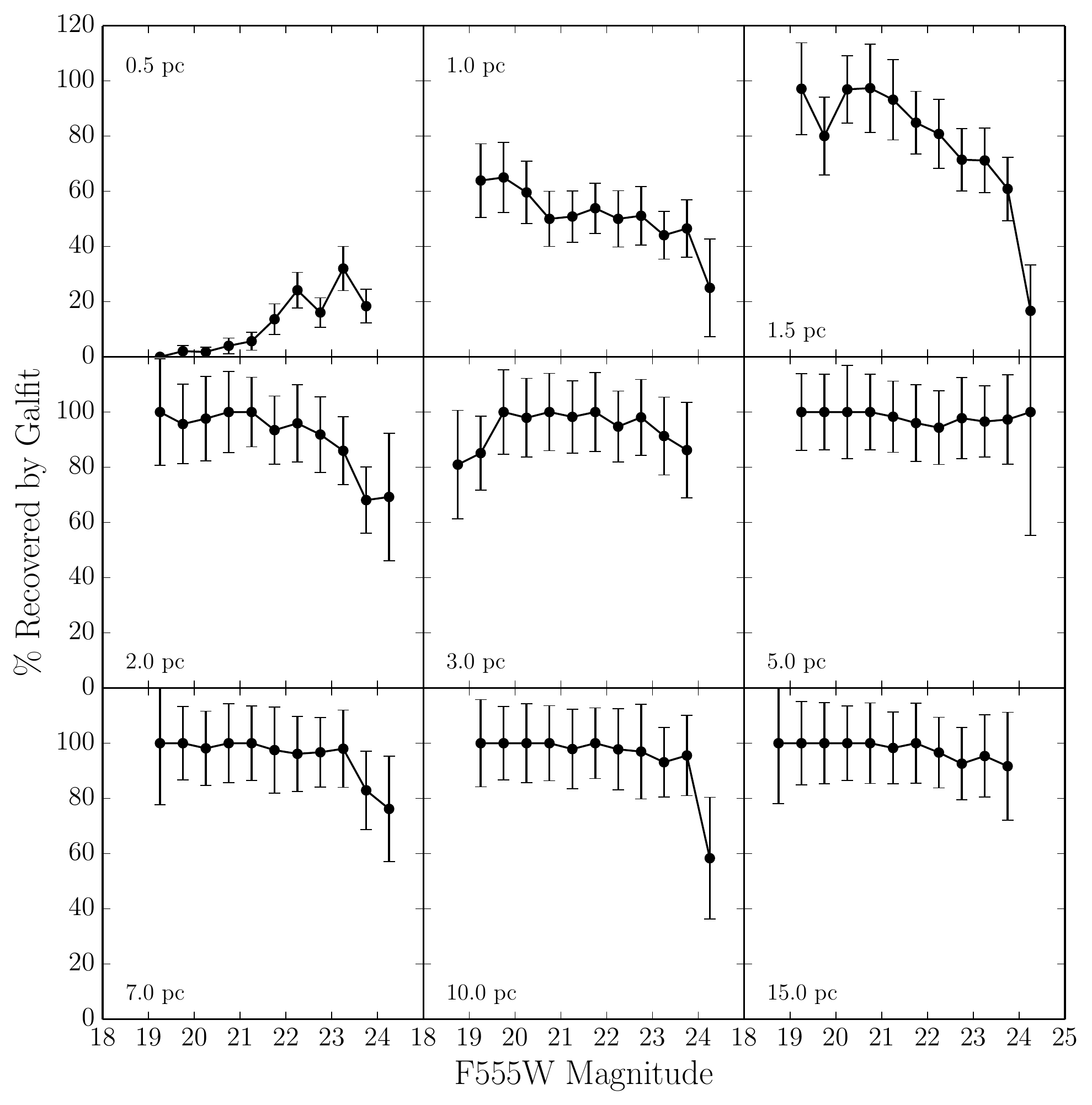}{\columnwidth}{(a) NGC 628c}
          \fig{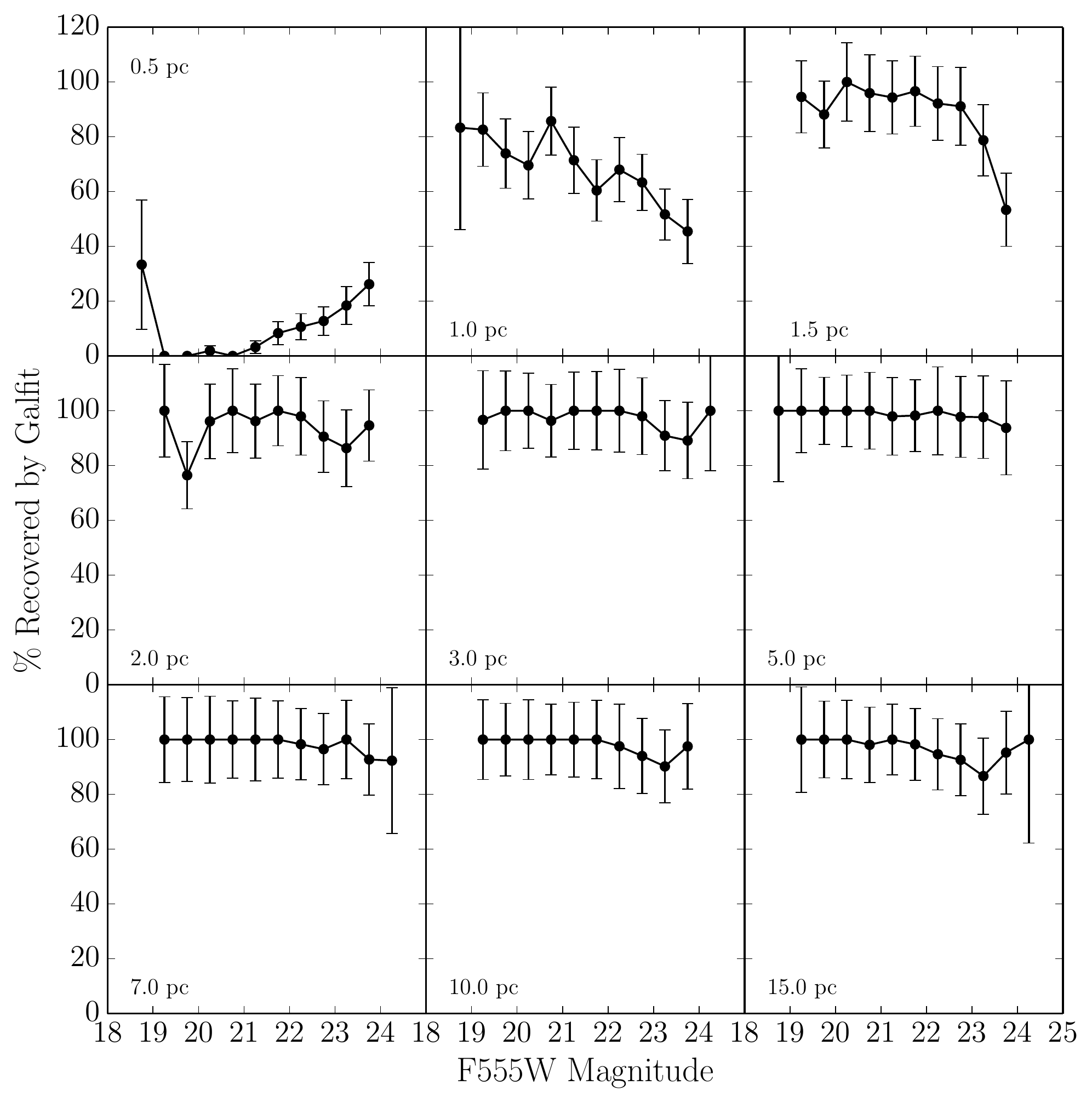}{\columnwidth}{(b) NGC 628e}}
\gridline{\fig{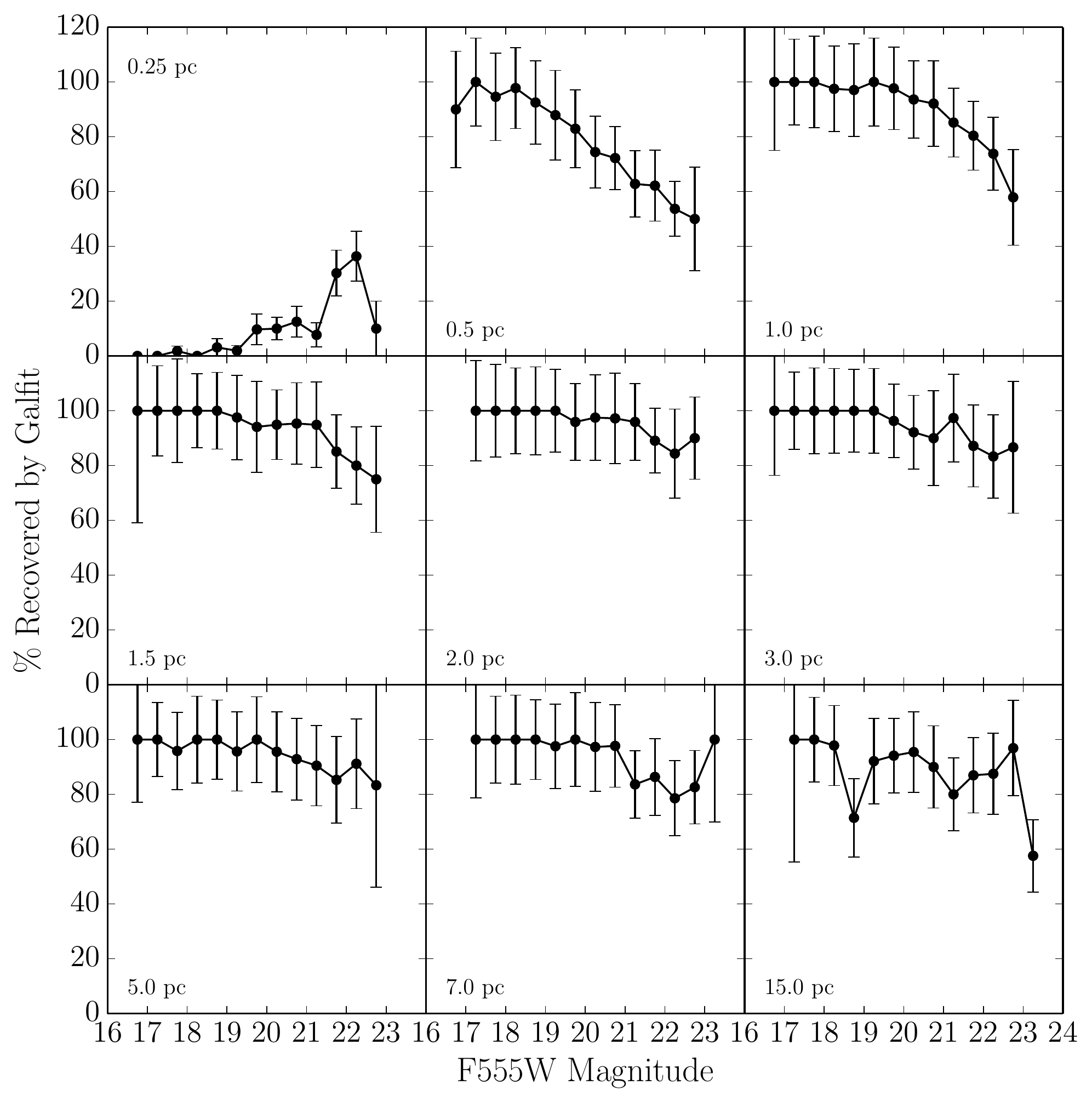}{\columnwidth}{(c) NGC 1313e}
          \fig{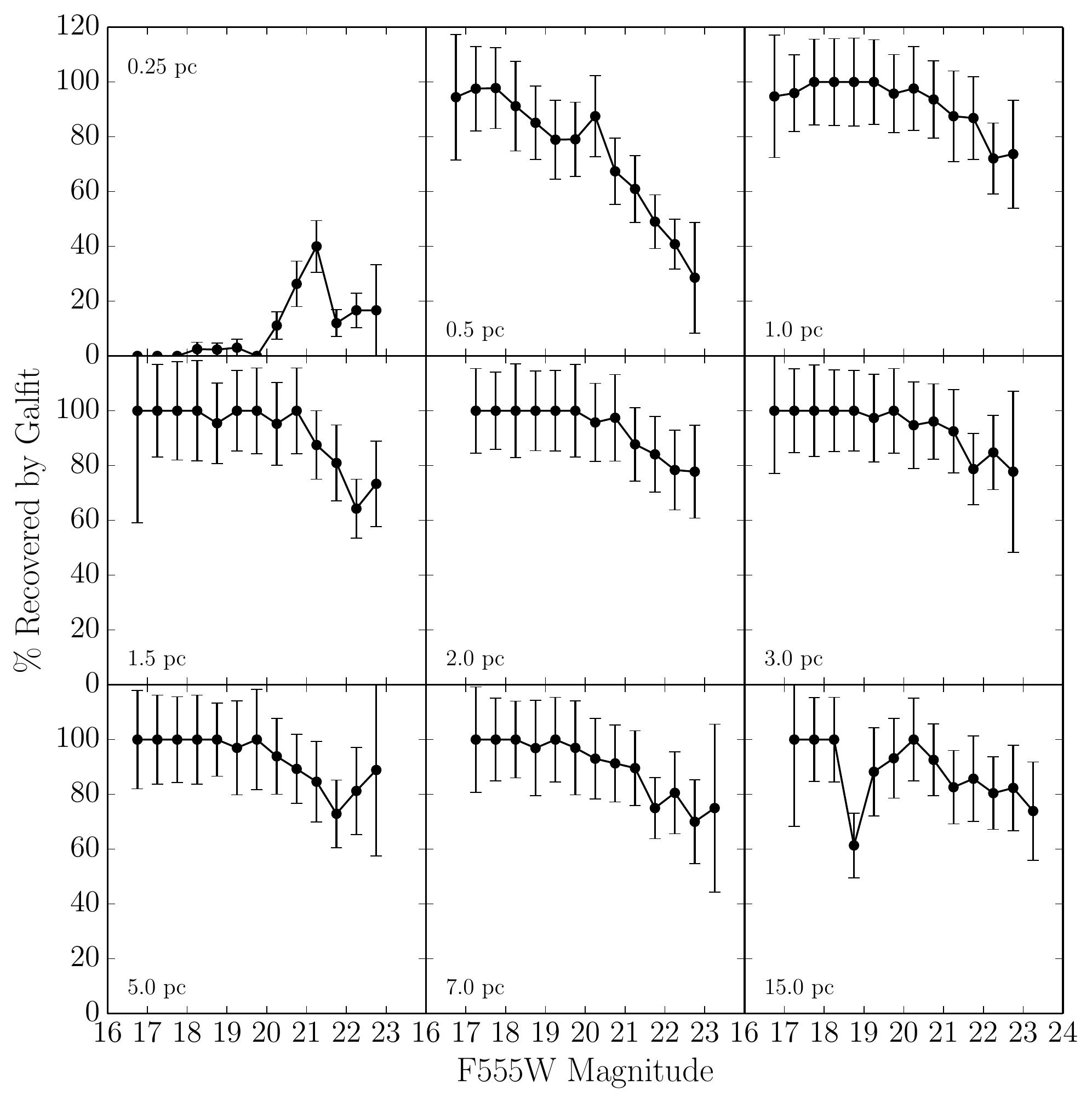}{\columnwidth}{(d) NGC 1313w}}
\caption{Percentage of recovered artificial clusters versus measured magnitude for each input effective radius for (a) NGC~628c, (b) NGC~628e, (c) NGC~1313e, and (d) NGC~1313w. The low recovered percentages for artificial clusters with sizes of $<$0.5~pc in NGC~628 and $<$0.25~pc in NGC~1313 are due to resolution effects (see text for details).
The error bars represent the Poissonian errors for the number of recovered clusters.
\label{recovery}}
\end{figure*}

We plot the results in Figure~\ref{recovery}. In NGC~628c and NGC~628e (Figure~\ref{recovery} (a) and (b), respectively), the majority of  clusters in the $r_{\mathrm{eff}}= 0.5$~pc bin are not recovered, probably because these objects straddle the boundary between unresolved stars and semi-resolved clusters. The recovery rate of clusters with 1.0~pc radii varies significantly over the magnitude range, between 30-70\% for NGC~628c, and 40-80\% for NGC~628e. For objects in the $r_{\mathrm{eff}}=1.5$~pc bin, the recovery percentage is 100\% at the bright end of the magnitude range, then begins to decrease to 50-60\% for both NGC~628 pointings, though the falloff occurs more rapidly for NGC~628c. However, the majority of 1.5~pc clusters in both pointings are recovered. For clusters with input $r_{\mathrm{eff}} \geq2.0$~pc, the recovery rate is essentially 100\% in all magnitude bins, except for some dips at the faint end in the larger radius panels. This indicates that GALFIT recovers almost all of the clusters greater than 2.0~pc in radius in both pointings, and the majority of sources between 1.0 and 2.0~pc in effective radius.

For each input radius, the recovery percentages are almost identical between the two NGC~1313 pointings (Figure~\ref{recovery} (c) and (d)). The majority of the 0.25~pc clusters are not recovered, again probably because they are close to the resolution limit between stars and clusters. The brightest objects in the 0.5~pc panels are 100\% recovered and followed by a falloff to 30-40\% recovered at the faintest magnitudes, but the majority of objects are recovered. Above 1.0~pc radii, essentially all objects are recovered, except for dips to $\sim$80\% at the fainter magnitudes. For NGC~1313, GALFIT is able to recover the majority of sources between 0.5~pc and 1.0~pc, and almost all sources larger than 1.0~pc in radius. The difference between recovery rates for artificial clusters with small radii for the two galaxies is likely due to the fact that NGC~628 is about twice as far away as NGC~1313.

\subsubsection{Bright Objects in LEGUS Stellar Catalogs}
\label{bright}

The artificial cluster tests in Section~\ref{artificial} assume that the sample input into GALFIT is not already biased against very compact clusters. To test this assumption, we search the final cluster catalog of each pointing for objects that satisfy $M_{V} \leq -6$, $M\geq5000$~M$_{\odot}$, age $\leq$ 200~Myr, and visual inspection class (mode) $\neq 1$ or 2. With this sample, we look for objects that may have been misclassified as stars (class 4) or associations (class 3) that could actually be very compact clusters that would fall in our parameter space. We also search the LEGUS version 1 stellar catalog (Sabbi et al. in prep.) for objects that satisfy $M_V \leq -8$. Very few stars are brighter than this absolute magnitude limit \citep{massey2006}, so any objects in the catalog brighter than this limit may be very compact star clusters.

We visually inspect these two samples of bright objects and remove spurious sources, including obviously saturated foreground stars, hot pixels near chip edges, and the nuclear star cluster. Any objects that overlap between the sample taken from the stellar catalog and the cluster catalog described in Section~\ref{obs-cat} are also removed. We then attempt to fit the remaining objects from both the cluster and stellar catalogs with GALFIT using the same input parameters as for the observed clusters (see Table~\ref{input}).

We inspect the residual images of objects that are recovered by GALFIT, meaning their fits successfully converged, and remove objects for which the fit was influenced by the presence of nearby objects. Of the remaining objects for NGC~628c, a total of 3 objects from the cluster catalog sample were measured to have small radii ($r_{\mathrm{eff}} \lesssim 3$~pc). From the stellar catalog, 5 objects had reasonable GALFIT fits with a radius $\lesssim$3~pc. For NGC~628e, only one object from the cluster catalog had a small radius, and none from the stellar catalog. 

For NGC~1313e, there were 6 objects from the cluster catalog and none from the stellar catalog with small radii. For NGC~1313w, there were 4 objects from the cluster catalog and none from the stellar catalog with small radii and reasonable GALFIT fits. These results indicate that few compact clusters are likely to have been excluded from the final LEGUS cluster catalogs for NGC~628 and especially for the nearer NGC~1313 system. In addition, given the results from Section~\ref{artificial}, the objects for which GALFIT fits were unsuccessful are unlikely to have $r_{\mathrm{eff}}$ between 1.0 and 3.0~pc in NGC~628, and 0.5 to 3.0~pc in NGC~1313.

\section{Results}
\label{results}

We present the structural parameters both measured with GALFIT and estimated from CI values for the cluster samples in NGC~628 and NGC~1313 in this section. Tables~\ref{results_628} and \ref{results_1313} contain our measurements for all clusters that satisfy the visual inspection class, age, and mass cuts described in Section~\ref{obs-cat}. Therefore, there are 320 clusters included in Table~\ref{results_628} for NGC~628 and 195 included in Table~\ref{results_1313} for NGC~1313. Those clusters that were successfully fit with GALFIT but for which $\eta \leq 1.3$ have very uncertain $r_{\mathrm{eff}}$ (as discussed in Section~\ref{reff}), and their $r_{\mathrm{eff}}$ values are enclosed by parentheses in the Tables. The astrophysical results that we present are based on these data, and thus are subject to inevitable observational selection effects.  However, the sample for this study was carefully selected to include star clusters that are likely to be bound and sufficiently massive so as to minimize the effects of stochastic sampling of their stellar populations. Thus this study includes what is perhaps the most complete sample of relatively massive (M $>$ 5000~M$_{\odot}$) young star clusters for which structural measurements have yet been obtained.

\begin{table*}
\centering
\caption{Properties of YMCs in NGC 628}
 \label{results_628}
\begin{tabular}{cccccccccc}
\hline
Cluster & R.A. & DEC & CI & $\eta$ & GALFIT & CI-estimated & $\log t_{\mathrm{age}}$ & $\log M$ & Mode \\
ID & (deg) & (deg) & (mag) & & $\log r_{\mathrm{eff}}$ (pc) & $\log r_{\mathrm{eff}}$ (pc) & (yr) & (M$_{\odot}$) & $f_{\mathrm{vis}}$ \\
\hline
196-c & 24.16912 & 15.80396 & 1.649 & \nodata & \nodata & 0.56$^{+0.10}_{-0.13}$ & 7.95$^{+0.35}_{-1.11}$ & 3.92$^{+0.22}_{-0.95}$ & 1.0 \\[1mm] 
237-c & 24.17594 & 15.80297 & 1.650 & 2.18 $\pm$ 0.19 & 0.38$^{+0.07}_{-0.08}$ & 0.56$^{+0.02}_{-0.02}$ & 7.04$^{+0.13}_{-0.00}$ & 4.11$^{+0.23}_{-0.00}$ & 1.0 \\[1mm] 
256-c & 24.17687 & 15.80270 & 1.491 & \nodata & \nodata & 0.29$^{+0.12}_{-0.17}$ & 8.30$^{+0.00}_{-0.30}$ & 3.72$^{+0.07}_{-0.16}$ & 1.0 \\[1mm] 
268-c & 24.16487 & 15.80224 & 1.664 & \nodata & \nodata & 0.58$^{+0.02}_{-0.02}$ & 7.48$^{+0.00}_{-0.00}$ & 4.37$^{+0.01}_{-0.03}$ & 2.0 \\[1mm] 
292-c & 24.17258 & 15.80185 & 1.722 & \nodata & \nodata & 0.65$^{+0.09}_{-0.12}$ & 8.00$^{+0.30}_{-1.15}$ & 3.93$^{+0.23}_{-0.94}$ & 2.0 \\[1mm]
\hline
\end{tabular}
\tablecomments{Col. (1): Cluster ID number. Objects located in the central pointing are designated ``-c'', while those in the east pointing are designated ``-e.'' Cols. (2) and (3): R.A. and Dec coordinates in decimal degrees (J2000). Col. (4): Concentration index. Col. (5): Power-law index, $\eta$, of the EFF light profile and the 1$\sigma$ error as reported by GALFIT. Col. (6): Log of the half-light (effective) radius in parsecs and the 1$\sigma$ positive and negative errors, measured by GALFIT. Parentheses denote objects best-described by $\eta< 1.3$. As discussed in Section~\ref{reff}, the GALFIT-determined effective radii of such clusters are not well-constrained and should be treated with caution. Col. (7): Log of the half-light (effective) radius in parsecs and the 1$\sigma$ positive and negative errors, estimated from the CI. Col. (8): Log of the best-fit cluster age in years and associated positive and negative errors allowed by the SED fits. Col. (9): Log of the best-fit cluster mass in solar masses and associated positive and negative errors allowed by the SED fits. Col. (10): LEGUS visual inspection class, mode. The full table is published in its entirety in the machine-readable format.  A portion is shown here for guidance regarding its form and content.\hfill
}
\end{table*}

\begin{table*}
\centering
\caption{Properties of YMCs in NGC 1313}
\label{results_1313}
\begin{tabular}{cccccccccc}
\hline
Cluster & R.A. & DEC & CI & $\eta$ & GALFIT & CI-estimated & $\log t_{\mathrm{age}}$ & $\log M$ & Mode \\
ID & (deg) & (deg) & (mag) & & $\log r_{\mathrm{eff}}$ (pc) & $\log r_{\mathrm{eff}}$ (pc) & (yr) & (M$_{\odot}$) & $f_{\mathrm{vis}}$ \\
\hline
45-e & 49.61172 & -66.45683 & 2.007 & 1.02 $\pm$ 0.05 & (8.00$^{+1.64}_{-8.00}$) & 0.67$^{+0.04}_{-0.04}$ & 8.30$^{+0.00}_{-0.00}$ & 4.72$^{+0.03}_{-0.00}$ & 1.0 \\[1mm] 
108-e & 49.64161 & -66.46315 & 1.871 & 1.01 $\pm$ 0.08 & (15.36$^{+2.44}_{-15.36}$) & 0.49$^{+0.05}_{-0.05}$ & 8.00$^{+0.00}_{-0.22}$ & 4.03$^{+0.04}_{-0.09}$ & 2.0 \\[1mm] 
133-e & 49.57730 & -66.46528 & 1.866 & 1.03 $\pm$ 0.07 & (5.16$^{+1.44}_{-5.16}$) & 0.48$^{+0.03}_{-0.03}$ & 8.30$^{+0.00}_{-0.00}$ & 3.99$^{+0.06}_{-0.04}$ & 2.0 \\[1mm] 
215-e & 49.60231 & -66.46820 & 1.652 & 1.16 $\pm$ 0.06 & (0.70$^{+0.24}_{-0.56}$) & 0.14$^{+0.02}_{-0.02}$ & 7.48$^{+0.00}_{-0.36}$ & 3.86$^{+0.05}_{-0.32}$ & 1.0 \\[1mm] 
228-e & 49.57860 & -66.46854 & 1.635 & 1.01 $\pm$ 0.03 & (14.74$^{+2.02}_{-14.74}$) & 0.11$^{+0.03}_{-0.03}$ & 8.30$^{+0.00}_{-0.00}$ & 3.98$^{+0.04}_{-0.04}$ & 2.0 \\[1mm]
\hline
\end{tabular}
\tablecomments{Col. (1): Cluster ID number. Objects located in the east pointing are designated ``-e'', while those in the west pointing are designated ``-w.'' Cols. (2) and (3): R.A. and Dec coordinates in decimal degrees (J2000). Col. (4): Concentration index. Col. (5): Power-law index, $\eta$, of the EFF light profile and the 1$\sigma$ error as reported by GALFIT. Col. (6): Log of the half-light (effective) radius in parsecs and the 1$\sigma$ positive and negative errors, measured by GALFIT. Parentheses denote objects best-described by $\eta< 1.3$. As discussed in Section~\ref{reff}, the GALFIT-determined effective radii of such clusters are not well-constrained and should be treated with caution. Col. (7): Log of the half-light (effective) radius in parsecs and the 1$\sigma$ positive and negative errors, estimated from the CI. Col. (8): Log of the best-fit cluster age in years and associated positive and negative errors allowed by the SED fits. Col. (9): Log of the best-fit cluster mass in solar masses and associated positive and negative errors allowed by the SED fits. Col. (10): LEGUS visual inspection class, mode. The full table is published in its entirety in the machine-readable format.  A portion is shown here for guidance regarding its form and content.\hfill
}
\end{table*}

\begin{table*}
\centering
\caption{Median Effective Radii}
\label{median-reff}
\begin{tabular}{ccccc}
\hline
LEGUS & GALFIT Median & CI Median & GALFIT & CI \\
Pointing & $r_{\mathrm{eff}}$ (pc) & $r_{\mathrm{eff}}$ (pc) & $N_{\mathrm{clusters}}$ & $N_{\mathrm{clusters}}$ \\
\hline
NGC~628c & 2.9 $\pm$ 0.9 & 3.2 $\pm$ 1.1 & 107 & 257 \\
NGC~628e & 2.9 $\pm$ 1.0 & 2.7 $\pm$ 0.9 & 27 & 63 \\ 
NGC~628 all & 2.9 $\pm$ 0.9 & 3.1 $\pm$ 1.1 & 134 & 320 \\
NGC~1313e & 2.7 $\pm$ 1.0 & 2.4 $\pm$ 0.9 & 14 & 54 \\
NGC~1313w & 2.0 $\pm$ 1.0 & 2.8 $\pm$ 1.2 & 45 & 141 \\
NGC~1313 all & 2.3 $\pm$ 1.2 & 2.7 $\pm$ 1.1 & 59 & 195 \\
\hline
\end{tabular}
\end{table*}

\subsection{Distribution of effective radii}
\label{reff}

As discussed in detail in Paper~I (see Section~4.1 and Figure~1 in that work), when GALFIT finds that a cluster's light profile slope is very shallow, $\eta \approx 1.0$, the effective radius we calculate is very uncertain. To avoid unphysical radii and errors, we restrict our analysis of the GALFIT samples to those objects with $\eta \geq 1.3$. The number of clusters remaining in the GALFIT samples after imposing this limit on $\eta$ is 134 in NGC~628 (107 in the central pointing and 27 in east pointing) and 59 in NGC~1313 (14 in the east pointing and 45 in west pointing). This cut on the GALFIT sample naturally introduces a bias against clusters with shallow radial intensity profiles which is why we also include results from the CI method. However, for clusters where we could obtain models with GALFIT, we expect the results to be more robust. 

\begin{figure*}
\centering
\includegraphics[width=0.6\textwidth]{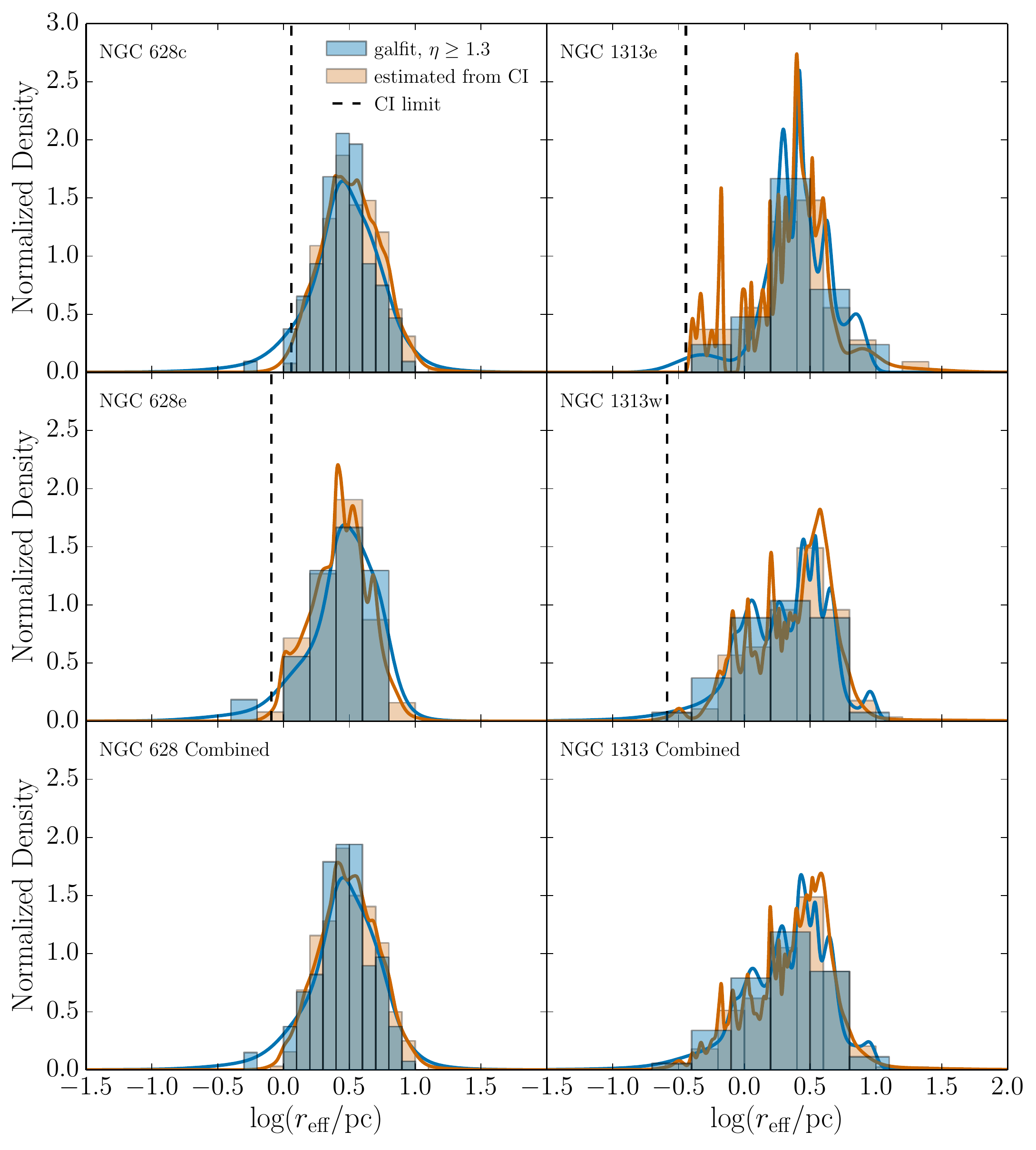}
\caption{Distribution of effective radii for NGC~628 (left column) and NGC~1313 (right column) clusters. The top two panels in each column show the radius distributions for the individual pointings, and the bottom panels show the pointings on each galaxy combined. Effective radii from GALFIT fits are plotted as blue histograms and CI-estimated effective radii are plotted as orange histograms. The solid curves show summed Gaussian kernels with widths equal to the error on each radius measurement. The vertical dashed lines show the location of the CI limit imposed on the LEGUS cluster catalog. \label{reff-hist}}
\end{figure*}

We plot the distribution of effective radii for NGC~628 and NGC~1313 in Figure~\ref{reff-hist}. The panels in the top and middle rows show the clusters located in the two pointings on each galaxy separately. The panels in the bottom row show the combination of the two pointings for each galaxy. The effective radii measured from GALFIT fits are plotted as blue histograms, while the CI-estimated effective radii are plotted as orange histograms. The width of the bins were estimated using the Freedman-Diaconis rule \citep{ivezic2014} and rounded to the nearest tenth for each panel. The blue and orange curves show the result of representing each cluster as a Gaussian kernel and summing the kernels together. 

Each Gaussian kernel is centered on the radius of a cluster, has a bandwidth (standard deviation) equal to the error on the radius of the cluster, and is normalized by the number of clusters in the sample.\footnote{This is a slight variation on the non-parametric statistical method known as kernel density estimation \citep{ivezic2014}, which is intended to give an accurate representation of the shape of the underlying distribution without using a histogram, which assumes bin width and placement. } The noisy appearance of these curves in the NGC~1313 panels is due to the relatively small effective radius errors for some of the clusters, which results in narrow Gaussian kernels.  We believe this may be the result of a much lower uncertainty in the distance to NGC~1313 as compared to NGC~628, which is combined with the FWHM and $\eta$ uncertainties to determine the effective radius errors. The CI limits listed in Table~\ref{catalog} for each pointing have been converted to effective radii using the relation between CI and effective radius determined in Section~\ref{ci-estimated}, and are plotted as vertical dashed lines.

In each individual panel, the distributions of GALFIT-measured radii and CI-estimated radii are very similar in shape, peak location, and overall extent. We perform Anderson-Darling tests to determine if the distributions in each panel are significantly different from each other. This takes the form of rejecting or accepting the null hypothesis, which is that the distributions in each panel are consistent with being drawn from the same parent distribution. In all panels except for NGC~628c, we cannot reject the null hypothesis ($p > 0.19$ for all five panels), meaning that the GALFIT-measured radii and CI-estimated radii are consistent with being drawn from the same parent population. For NGC~628c, we can reject the null hypothesis at a significance level of 7\% ($p = 0.07$), suggesting that the two distributions in this panel are somewhat different from each other. This probably is due to the offset of the CI sample towards larger radii by a small but statistically significant amount.

The median effective radii are listed in Table~\ref{median-reff} for each pointing separately and for the combined sample from each galaxy. We find that the two methods for estimating effective radii produce very similar medians and MADs, which gives confidence in the CI-estimated effective radius technique. This technique can therefore be used to derive effective radii from much larger samples of clusters than is possible with our GALFIT method.

Comparing NGC~628 to NGC~1313, the overall shape of the effective radius distributions are relatively similar, and also resemble the distribution presented in Paper~I for M83 (see Figure~2 in that work). To a first approximation the distributions of r$_{\rm eff}$ in M83, NGC~628, and NGC~1313 have lognormal shapes and peaks at $\sim$3~pc. We perform Anderson-Darling tests to statistically compare the GALFIT-measured and CI-estimated radius distributions for NGC~628 and NGC~1313, and find that we can reject the null hypothesis at high significance for both methods of measurement ($p=0.001$ for GALFIT and $p=1.3\times10^{-5}$ for CI). Therefore, despite their approximate similarities, the distributions of effective radii in these two galaxies are formally inconsistent with being drawn from the same parent population. 

We suggest that this difference is an observational artifact because in NGC~628, the smallest cluster radius we measure is about 1~pc, whereas in NGC~1313 (and M83 as well), we find a tail in the distribution to small radii, $\sim$0.3~pc. Because NGC~628 is about twice as distant as NGC~1313 and M83, the CI limit for sample selection corresponds to a larger radius, and therefore likely results in the smallest clusters being removed from that sample. From Figure~\ref{recovery}, at least 40\% of artificial clusters with a radius of 1~pc are lost at all magnitudes in NGC~628 whereas for NGC~1313, all input clusters at this radius are well recovered. Therefore, we conclude that the physical distributions of cluster $r_{\rm eff}$ are similar to each other, at least when considering the parameter space covered by this study.

In all three galaxies, the largest clusters for which $\eta \geq 1.3$ are about 10~pc in radius. We do not expect to find very large clusters in the GALFIT-measured samples with this conservative $\eta$ cut, and we also do not find very large clusters in the CI-estimated radius distributions in Figure~\ref{reff-hist}. However, we note that the CI method of estimating $r_{\mathrm{eff}}$ becomes less sensitive as the CI increases, that is, a small uncertainty in CI leads to a big uncertainty in $r_{\mathrm{eff}}$ for large $r_{\mathrm{eff}}$. An extended tail of clusters with large radii appears in the M83 distribution when clusters with shallower light profiles are included ($\eta \geq 1.1$, see bottom panel of Figure~2 in Paper~I). Therefore, the properties and numbers of clusters with shallow profiles remains uncertain.

Several studies of YMCs and GCs in the Milky Way and other nearby galaxies have found strikingly similar effective radius distributions to those we present here and in Paper~I \citep[e.g.,][]{larsen2004,scheepmaker2007,barmby2009,bastian2012a,puzia2014}. In particular, the location of the peak in the observed distribution of r$_{\rm eff}$ appears to be quite robust, and is consistently found to be at 2 to 3~pc across a range of cluster mass, age, and environment. We find that NGC~628 and NGC~1313 provide more evidence in support of this conclusion, especially given the results of our completeness tests in Section~\ref{completeness} and the fact that the CI limits are located significantly below the peaks of the effective radius distributions.

\subsection{Effective radius as a function of cluster properties}
\label{reff-properties}

\begin{figure*}
\centering
\includegraphics[width=0.6\textwidth]{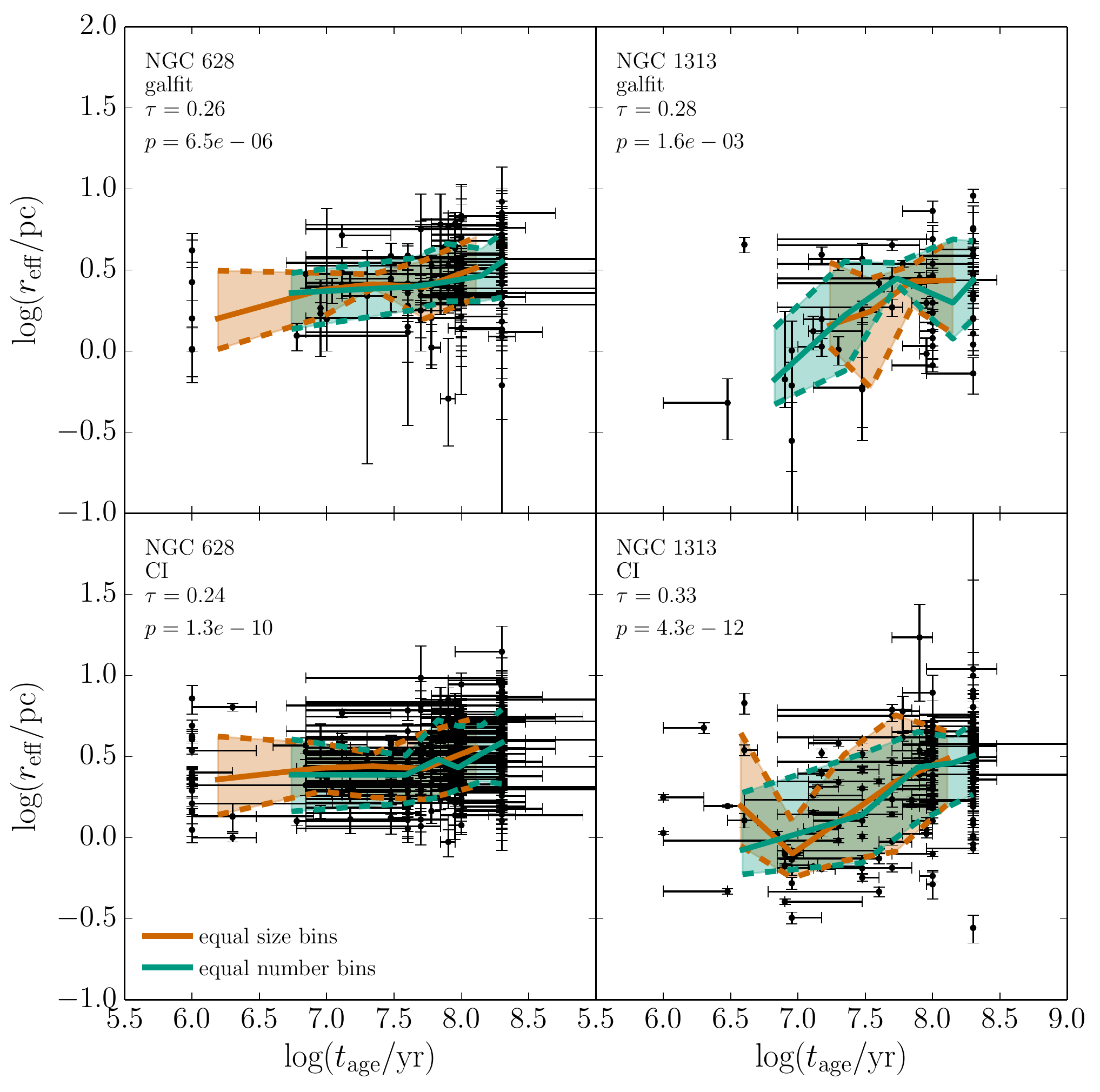}
\caption{Effective radius as a function of cluster age for NGC~628 (left column) and NGC~1313 (right column). The top row of panels show the effective radii measured with GALFIT, and the bottom row show those estimated from CI values. The solid lines show median effective radii in equal size bins in log-space (orange) and in bins with equal numbers of clusters (green). The dashed lines and shaded areas show the extent of the 16th and 84th percentiles in radius for each binning method. The Kendall $\tau$ correlation statistic and associated $p$-value are located in the top left corner of each panel. \label{reff-age} }
\end{figure*}

In Figure~\ref{reff-age}, we plot effective radius as a function of age for the clusters in NGC~628 (left column) and NGC~1313 (right column). We combine the cluster samples from the individual pointings into one for each galaxy since their properties were not found to be significantly different in Figure~\ref{reff-hist}. The top panels show the effective radii measured with GALFIT, and the bottom panels show the CI-estimated effective radii. In each panel, we plot the median effective radii in equal size bins in $\log(t_{\mathrm{age}})$-space (orange lines) and in bins containing equal numbers of clusters (green lines). Each bin contains at least 5 clusters. The dashed lines of each color show the 16th and 84th percentiles in radius for each bin. 

We calculate Kendall's $\tau$, a nonparametric correlation coefficient, to determine if the effective radii and cluster ages are correlated in each panel (see Appendix~\ref{appendix} for a short discussion of the definition and applicability of Kendall's $\tau$). We find modest correlations in both panels for NGC~628 with high significance. As labeled in the figure, Kendall's $\tau$ statistic is $\sim$0.25 for both panels, and the associated $p$-values are $p = 6.5\times10^{-6}$ for the GALFIT sample and $p=1.3\times10^{-10}$ for the CI sample. Because these $p$-values are smaller than our selected significance level of $p=2.7\times10^{-3}$ (corresponding to a 3$\sigma$ significance level for a two-tailed test), we can reject the null hypothesis of the Kendall's $\tau$ correlation test, which means we confirm statistically significant correlations for both panels.

\begin{figure*}
\centering
\includegraphics[width=0.6\textwidth]{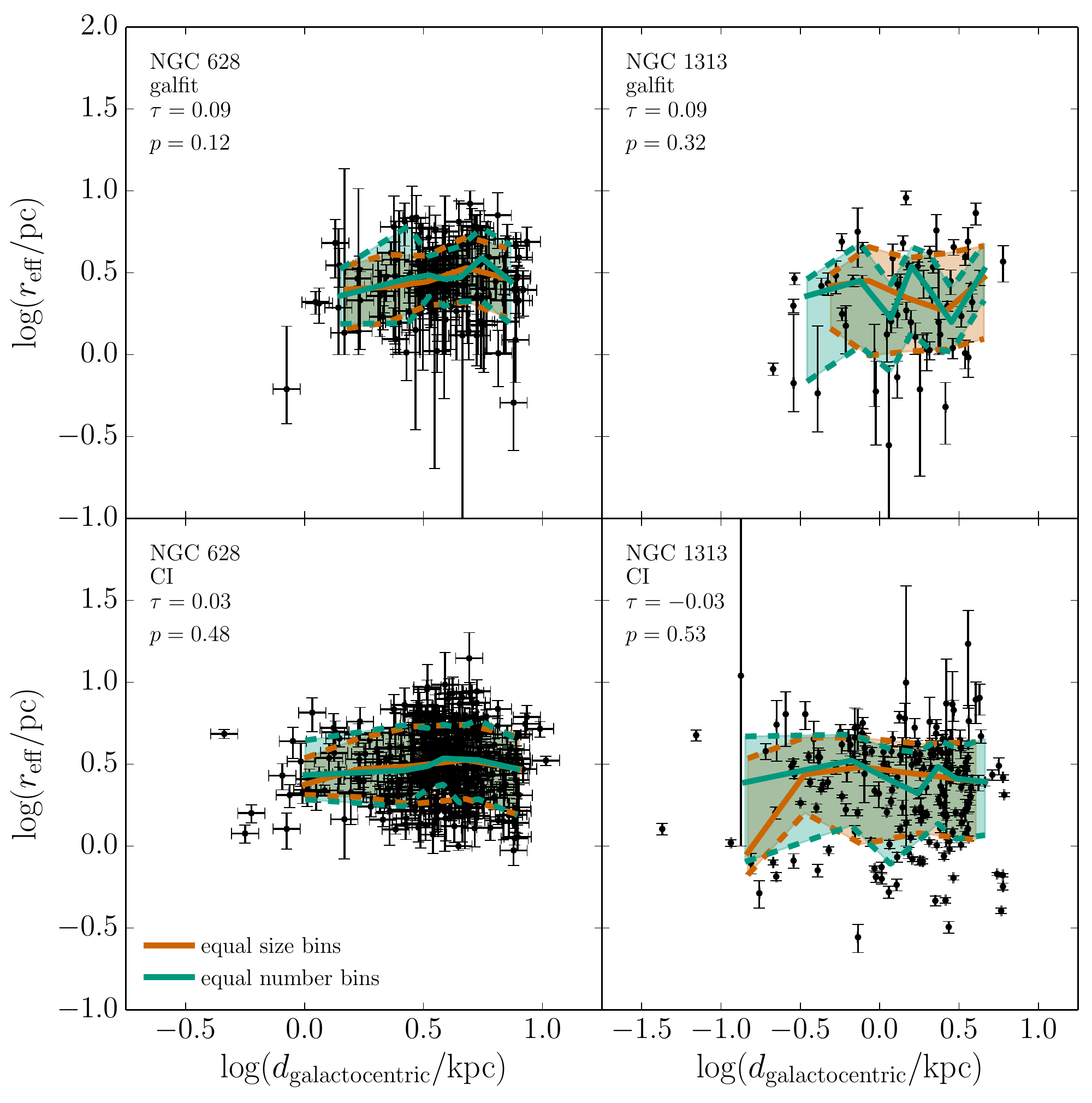}
\caption{Effective radius as a function of galactocentric distance for NGC~628 (left column) and NGC~1313 (right column). The effective radii measured with GALFIT and estimated from CI values are located in the top and bottom row of panels, respectively. The solid lines, dashed lines, and shaded regions are the same as in Figure~\ref{reff-age}. The Kendall $\tau$ correlation statistic and associated $p$-value is given in each panel. Note the different scales on the x-axis for NGC~628 versus NGC~1313. \label{reff-dist} }
\end{figure*}

Assuming a power-law relation exists between cluster radius and age, we perform simple least-squares fits to the median effective radii from each binning procedure, including standard errors on the medians, and find very shallow slopes of $\sim$0.1. For NGC~1313, we find slightly stronger correlations, also with high significance. For the GALFIT sample, $\tau \sim 0.28$ and $p=1.6\times10^{-3}$, while $\tau \sim 0.33$ and $p=4.3\times10^{-12}$ for the CI-estimated sample. The least-squares fits result in slopes of $\sim$0.3, which agrees with the trend found in Figure~3 in Paper~I.

We note that the smaller sample size for NGC~1313 may lead to larger scatter and increased uncertainty in the radius-age relation than for NGC~628. However, the cluster samples in NGC~1313 and M83 extend to smaller radii than NGC~628, and the strength of the radius-age relation for these two galaxies is greater than for NGC~628. The proximity of both NGC~1313 and M83 ($\sim$4~Mpc) has allowed us to detect smaller clusters in those galaxies as compared to NGC~628, and perhaps has provided better leverage on the radius-age relation. As discussed in Paper~I, a few studies have found slight positive or negative correlations between radius and age, while others have found essentially no correlation at all \citep[e.g.,][]{larsen2004, mclaughlin2005, scheepmaker2007, bastian2012a}. The positive correlations presented here and in Paper~I suggest that on average, YMCs expand slightly over the first few hundred Myrs of evolution. 

In Figure~\ref{reff-dist}, we plot effective radius as a function of distance from the center of the galaxy for the clusters in NGC~628 (left column) and NGC~1313 (right column). Again we combine the cluster samples from the individual pointings for each galaxy, and the top panels show the radii measured with GALFIT, while the bottom panels show CI-estimated radii. The solid and dashed lines are the same as in Figure~\ref{reff-age}, and show running medians and 1$\sigma$ percentiles in effective radius for different binning techniques. 
\begin{figure*}
\centering
\includegraphics[width=0.6\textwidth]{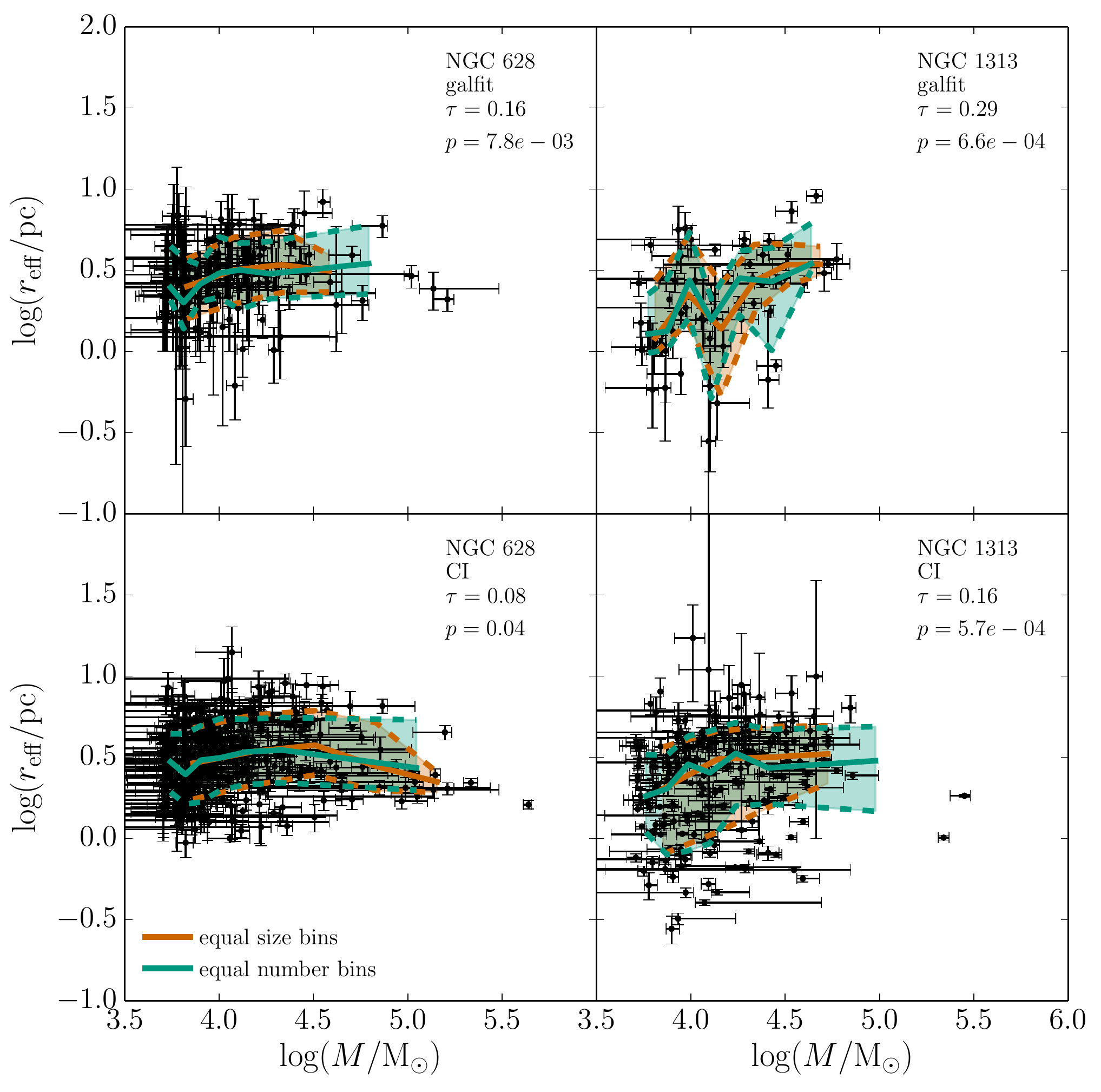}
\caption{Effective radius as a function of cluster mass for NGC~628 (left column) and NGC~1313 (right column). The top row of panels show the effective radii measured with GALFIT, and the bottom row show those estimated from CI values. The solid lines, dashed lines, and shaded regions are the same as in Figure~\ref{reff-age}. The Kendall $\tau$ correlation statistic and associated $p$-value is given in each panel. \label{reff-mass} }
\end{figure*}

The clusters in NGC~628 are located between $\sim$300~pc and 10~kpc from its center, while the clusters in NGC~1313 extend closer to the galactic center, the closest being about 30~pc from the center. No correlation between cluster radius and galactocentric distance is apparent to the eye. We again calculate Kendall's $\tau$ to determine if a correlation is present, and find no statistically significant correlations in any of the four panels of Figure~\ref{reff-dist}. We divide the samples in both galaxies into three bins in cluster age, $<$50~Myr, 50-100~Myr, and 100-200~Myr, chosen to reflect the age ranges over which the mean cluster mass is relatively constant (especially for NGC~628) as shown in Figure~\ref{age-mass}. Dividing the samples into bins in age also minimizes any effect of cluster expansion with time. We again find no statistically significant correlations with the Kendall's $\tau$ correlation test in any of the bins in cluster age. 

We are able to probe a larger range of galactocentric distance with these data than with the M83 sample, but no significant correlation is found in any of the three galaxies. Again, the evidence from other studies is mixed; some find slight correlations between YMC radius and distance, and others find none \citep[e.g.,][]{scheepmaker2007,bastian2005,barmby2009,bastian2012a,Sun16}. The sizes of globular clusters appear to be shallowly related to their galactocentric distances, though in some cases, like the Milky Way, the relationship is stronger \citep[e.g.][]{vandenbergh1991,harris2009, harris2010, masters2010}. 

 Models where young star clusters quickly fill their tidal radii are attractive. But in this case the radii of star clusters should eventually reflect their local tidal gravitational fields that usually decrease with increasing galactocentric radius. Our results suggest that clusters across the range of mass, age, and environment covered by our samples do not show effects of tidal truncations in their half-light radii, otherwise, a stronger increase of cluster size with galactocentric distance is expected \citep[e.g.][]{gieles2011b,madrid2012,webb2016}.

In Figure~\ref{reff-mass}, we plot effective radius as a function of cluster mass for the clusters in NGC~628 (left column) and NGC~1313 (right column). Again, the cluster samples from the individual pointings on each galaxy are combined, and the top panels show the radii measured with GALFIT, while the bottom panels show CI-estimated radii. The solid and dashed lines are the same as in Figure~\ref{reff-age}. No strong correlation is obvious in the NGC~628 panels, and none is present, according to the Kendall's $\tau$ correlation test. In NGC~1313, however, we find statistically significant trends in both the GALFIT ($\tau= 0.3$ and $p=6.6\times10^{-4}$) and CI-based samples ($\tau= 0.2$ and $p=5.7\times10^{-4}$). This may be another indication of the importance of including clusters with small r$_{\rm eff}$, which tend to be young and have lower masses. Assuming a power-law relation between cluster radius and mass, we perform simple least-squares fits to the median effective radii from both binning techniques including standard errors on the medians. We find steeper power-law slopes for the GALFIT sample, $\sim$0.5, than for the CI-estimated sample, $\sim$0.2 to 0.3.

\begin{figure*}
\centering
\includegraphics[width=0.6\textwidth]{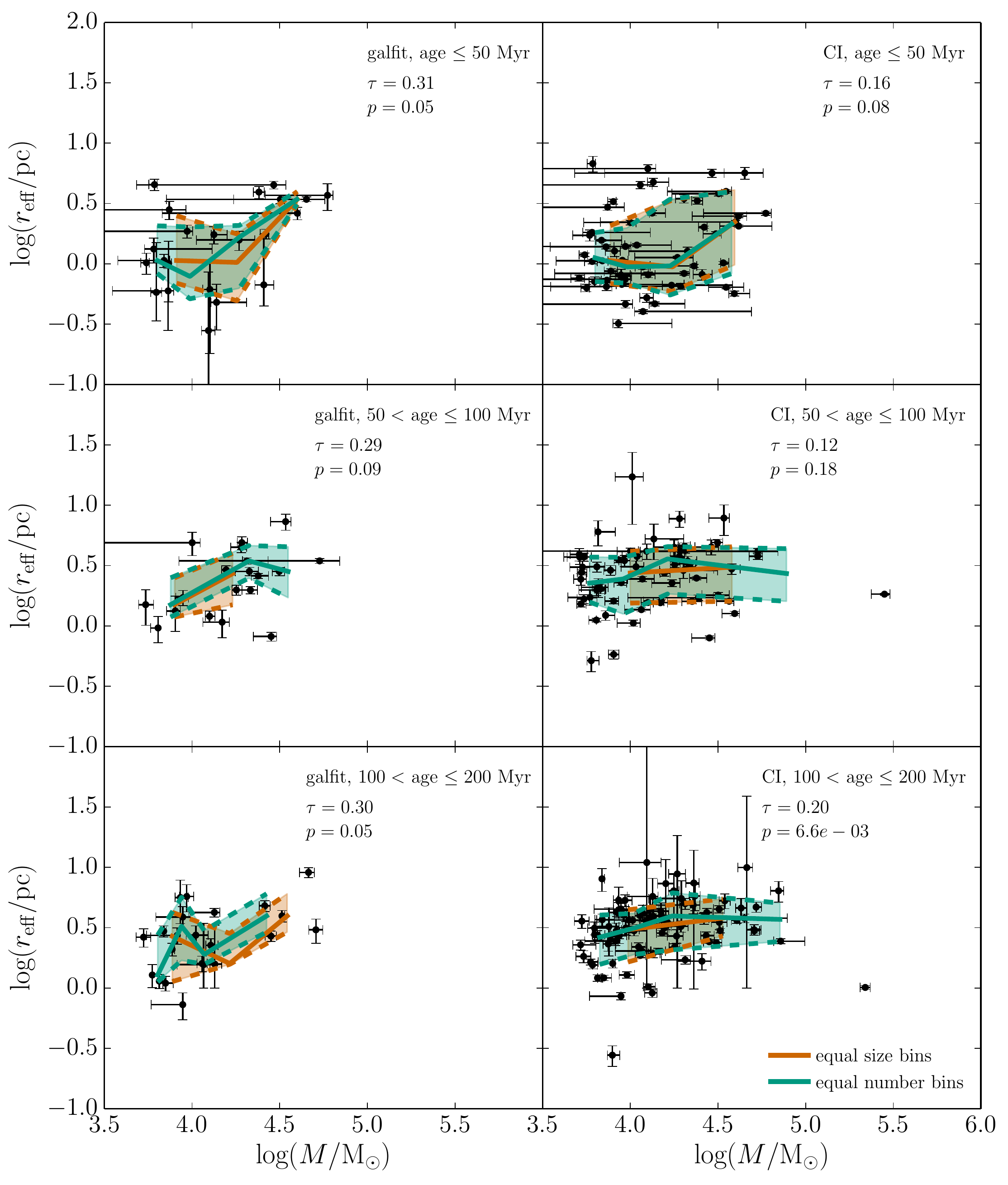}
\caption{Effective radius as a function of cluster mass for NGC~1313 in three bins in cluster age, 0-50~Myr, 50-100~Myr, and 100-200~Myr. The left column shows the effective radii measured with GALFIT, and the right column shows those estimated from CI values. The solid lines, dashed lines, and shaded regions are the same as in Figure~\ref{reff-age}. The Kendall $\tau$ correlation statistic and associated $p$-value is given in each panel. \label{reff-mass-agebins} }
\end{figure*}

We further investigate the correlation found in Figure~\ref{reff-mass} for NGC 1313 by dividing the sample into three bins in cluster age ($<$50~Myr, 50-100~Myr, and 100-200~Myr) and plotting the results in Figure~\ref{reff-mass-agebins}. Again, the GALFIT sample is plotted in the left column and the CI-estimated sample in the right column. Although correlations are apparent to the eye, by calculating Kendall's $\tau$ for each panel, we find no significant correlations between effective radius and cluster mass. Perhaps the radius-age relation for NGC~1313 drives the correlation found in Figure~\ref{reff-mass} because more massive clusters have a higher likelihood of being older, statisically (see Figure~\ref{age-mass}). Alternatively, the small numbers of clusters in each age bin may prevent a statistically significant correlation from being found. In addition, it is possible our cluster samples do not extend to old enough ages to detect a correlation between effective radius and cluster mass, as found in the oldest bin in Paper~I.

Previous studies of clusters below $\sim$10$^6$~M$_{\odot}$ have found little to no correlation between effective radius and cluster mass \citep[e.g.,][]{larsen2004,bastian2005, scheepmaker2007, barmby2009, bastian2012a}. Above $\sim$10$^6$~M$_{\odot}$, the radii of both YMCs and globular clusters appear to increase with mass (or luminosity) \citep[e.g.,][]{kisslerpatig2006,harris2009, fall2012,bastian2013}.

\subsection{EFF Profile Indices}
\label{eff-index}

The power-law index of the EFF light profile, $\eta$, describes the slope of the wings of a star cluster's intensity profile. We find that the range of light profile indices of the clusters fit by GALFIT are $1.0 \leq \eta \leq 9.3$ in NGC~628, and $1.0 \leq \eta \leq 5.6$ in NGC~1313, though the majority of clusters in both galaxies are best described by $1.0 \leq \eta \leq 3.0$. The median $\eta$ values are 1.4 for NGC~628 and 1.2 for NGC~1313. The smallest possible value of $\eta$ is artificially set to 1.0 by GALFIT because a light profile with $\eta < 1.0$ must, by definition, contain an infinite amount of light. However, $\eta$ values smaller than 1.0 have been recovered for other clusters using different techniques \citep{larsen2004}. The median $\eta$ values we find could therefore be smaller in reality. In any case, the range of $\eta$ values we find agrees with those presented in Paper~I and several other studies of cluster radial intensity profiles in nearby galaxies \citep[e.g.][]{elson1987, mackey2003a, glatt2009}. 

We also look for any relationships between $\eta$ and cluster age, mass, and galactocentric distance, similarly to Section~\ref{reff-properties}. We perform Kendall's $\tau$ correlation tests and find no statistically significant correlations between $\eta$ and any cluster property. This agrees with our findings for M83 in Paper~I. Therefore, we do not find evidence to support the increase in $\eta$ with age as reported by \cite{larsen2004}, which may indicate that the light profiles of YMCs are rather robust during their early evolution.

\subsection{Dynamical Age}
\label{dynamical-age}

We also calculate the dynamical age, $\Pi$, as defined by \cite{gieles2011a}, for our objects. $\Pi$ is the ratio of cluster age to the crossing time, $\Pi \equiv t_{\mathrm{age}}/t_{\mathrm{cross}}$, and can be used to determine if a system is gravitationally bound at the present time. We use Eq. 1 from \cite{gieles2011a} to calculate crossing times with the measured effective radii and masses. If the age of a stellar system exceeds its current crossing time, $\Pi > 1$, then the stars in that system have remained clustered together in space for their lifetimes, and have not freely expanded into their surroundings, which implies that the system is likely gravitationally bound. As discussed in \cite{gieles2011a}, the assumption of virial equilibrium in Eq. 1 overestimates the crossing time of unbound associations as the objects age and expand freely. Therefore, for older associations, $\Pi$ is underestimated, aiding in the distinction between bound and unbound objects. Of course, the true dynamical state of a stellar system cannot be determined without complete kinematical measurements of the constituent stars, but such measurements are non-trivial and beyond the scope of this work. Despite this limitation, \cite{portegieszwart2010} and \cite{bastian2012a} have found the dynamical age, $\Pi$, to be a useful discriminator between bound clusters and unbound stellar associations. 

\begin{figure}
\centering
\includegraphics[width=\columnwidth]{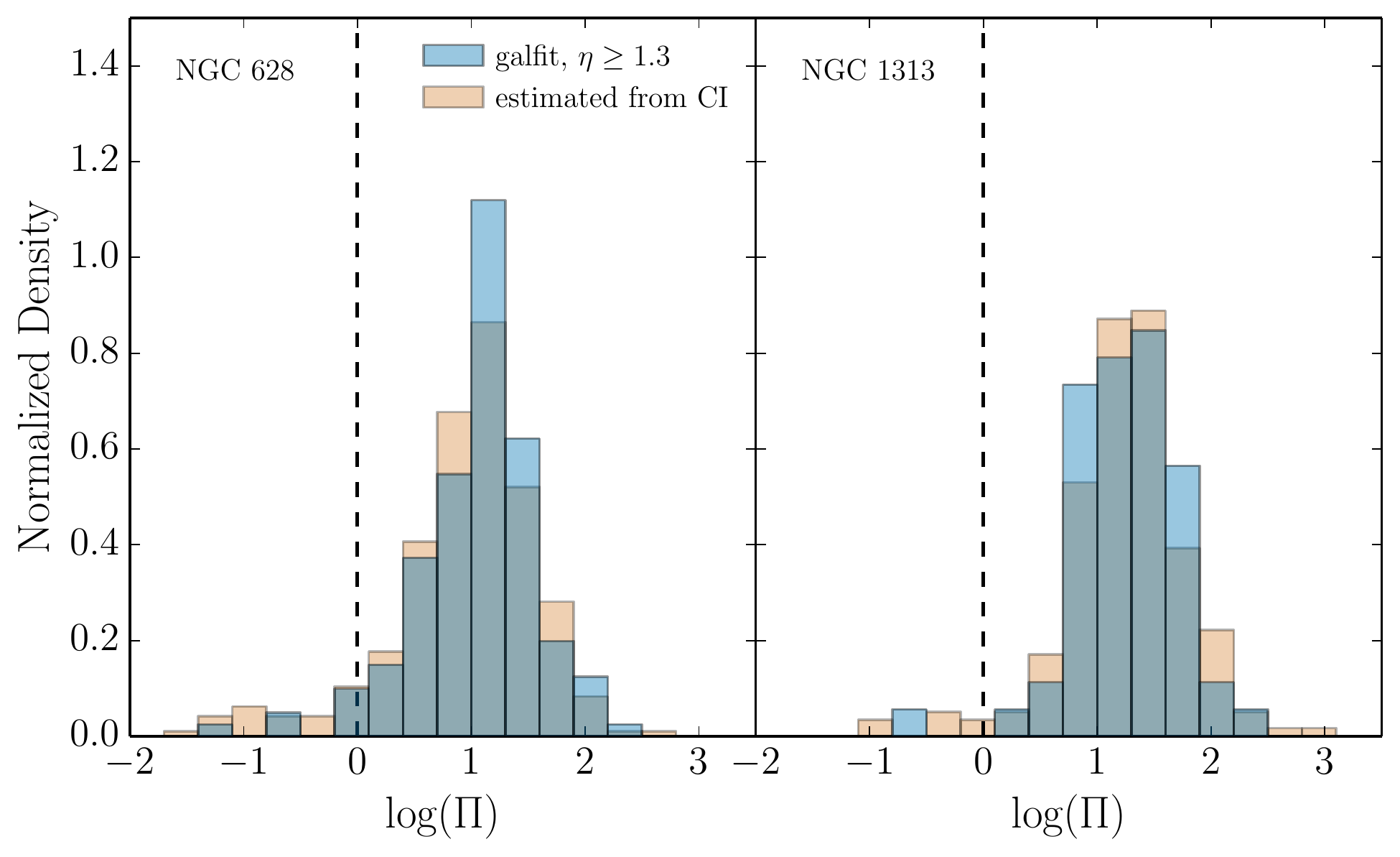}
\caption{Distribution of dynamical age, $\Pi$, of clusters in NGC~628 (left) and NGC~1313 (right). Dynamical ages calculated using effective radii from GALFIT fits are plotted as blue histograms and those using CI-estimated effective radii are plotted as orange histograms. The vertical dashed lines separate the bound ($\log(\Pi) > 0$) and unbound ($\log(\Pi) < 0$) objects.\label{pi-hist} }
\end{figure}

In Figure~\ref{pi-hist}, we present the distribution of dynamical ages of the clusters in NGC~628 (left panel) and NGC~1313 (right panel). The blue (darker) histograms correspond to dynamical ages calculated with effective radii determined from GALFIT, and the orange (lighter) histograms correspond to those using CI-estimated effective radii. The black dashed lines separate the bound and unbound objects, according to $\Pi$. The peak in the $\Pi$ distribution (particularly for NGC 628) is a selection effect, and is a consequence of the clusters have a roughly constant radius and an age limit of 200~Myr.

Clearly, the vast majority of objects in both galaxies fall in the region $\Pi > 1$, which means they are most likely bound systems. In NGC~628, only 4\% of objects in the GALFIT sample and 8\% of the objects in the CI-estimated sample can be classified as unbound. In NGC~1313, those percentages are 4\% and 2\%, respectively. Essentially all of the objects for which $\Pi < 1$ are quite young, $\lesssim10$~Myr, so the dynamical state of these few objects is ambiguous, as noted in \cite{gieles2011a}. However, it appears that the LEGUS selection criteria and visual inspection techniques successfully produce samples of (likely) gravitationally bound star clusters, as intended \citep{adamo2017}.

\subsection{Jacobi Radii}
\label{jacobi-radii}

The Jacobi radius of a cluster can help to determine if the tidal field of the host galaxy has a strong influence on the evolution of the cluster. It sets the size of the zero-velocity surface of a cluster in a tidal field, and therefore also defines the volume over which stars are bound to the cluster. If the ratio of the half-mass radius (about $4/3\times r_{\mathrm{eff}}$) to the Jacobi radius is about 0.15 or 0.2, the cluster is filling its Roche volume, and its evolution will be affected by the galaxy's tidal field \citep{henon1961,alexander2014}. A cluster with a half-mass to Jacobi radius ratio less than 0.15 or 0.2 will evolve as essentially an isolated system and expand gradually. A tidally-filling cluster should contract as it loses mass \citep{heggiehut2003}.

For a galaxy with a flat rotation curve and a non-rotating cluster, the Jacobi radius is defined by Equation~9 in \cite{portegieszwart2010},
\begin{equation}
r_{\mathrm{J}} = \left(\frac{GM}{2\omega^2}\right)^{1/3} \label{jacobi}
\end{equation}
which requires knowledge of the cluster mass, $M$, and angular speed of the galaxy's rotation, $\omega$.  According to H$\alpha$ kinematics from \cite{daigle2006}, NGC~628 has a flat rotation curve with a de-projected rotation velocity of about 175~km/s beyond $\sim$2.4~kpc in the galaxy's disk. NGC~1313, on the other hand, has an H\,{\sc i} rotation curve that is increasing linearly within $\sim$7.5~kpc, and the behavior beyond this distance is unclear \citep{ryder1995}. Because Equation~\ref{jacobi} assumes a flat rotation curve, we can only estimate Jacobi radii for clusters in NGC~628 at this time. There are few observational studies of the rotation of young clusters themselves, and those that do exist find that the rotational energy of such objects is relatively small \citep[e.g.,][]{henaultbrunet2012}.

\begin{figure}
\centering
\includegraphics[width=\columnwidth]{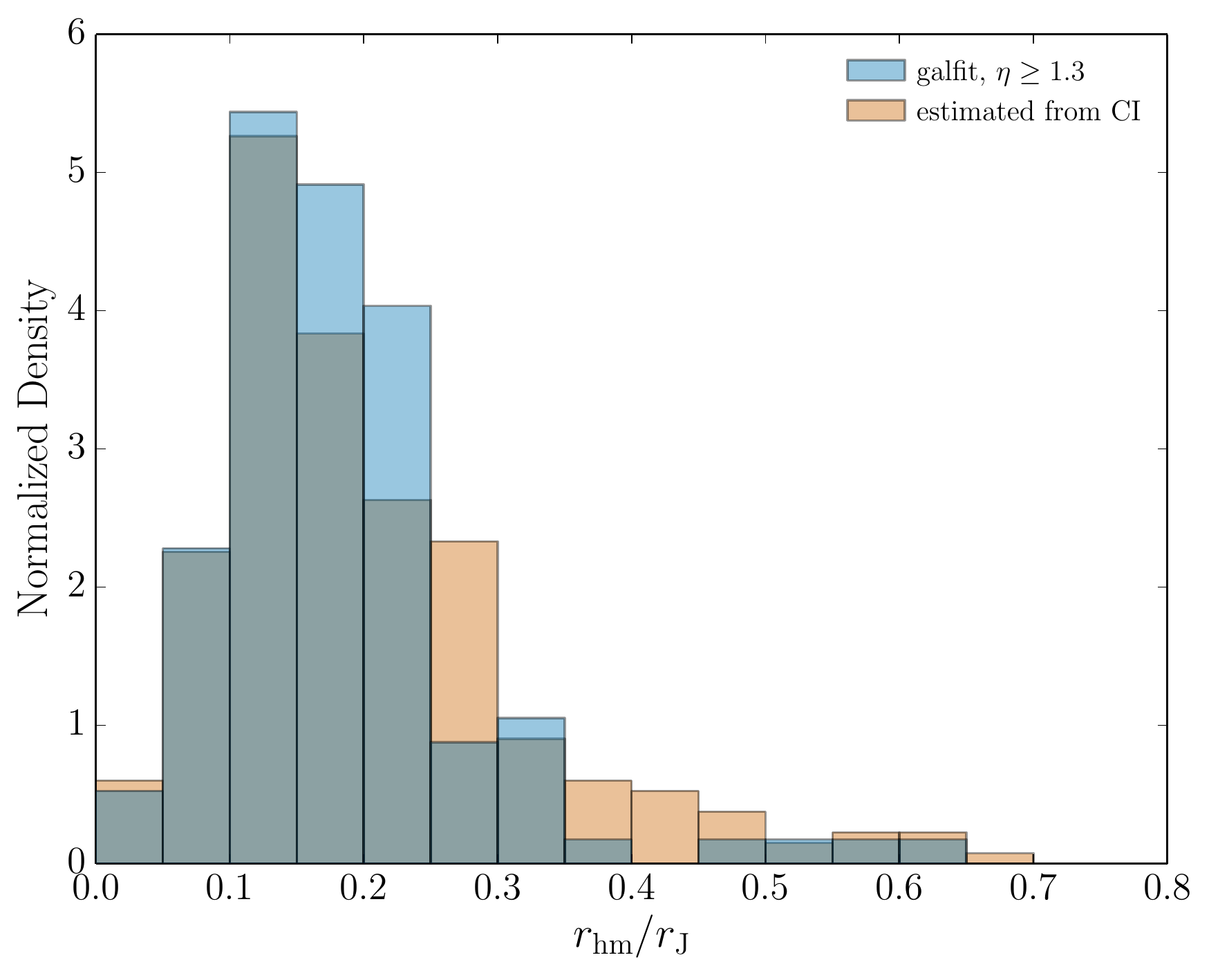}
\caption{Distribution of the ratio of half-mass radii ($r_{\mathrm{hm}}$) to Jacobi radii ($r_{\mathrm{J}}$) for clusters in NGC~628. Ratios calculated using effective radii from GALFIT fits are plotted as a blue histogram and those using CI-estimated effective radii are plotted as an orange histogram. \label{jacobi-ratio} }
\end{figure}

For each cluster located at a galactocentric distance $>2.4$~kpc in NGC~628, we estimate the angular speed of the cluster from the rotation velocity of the disk and the galactocentric distance of the cluster, $\omega = v_{\mathrm{rot}}/d_{\mathrm{gc}}$. We find Jacobi radii of 14 to 61~pc for clusters in the GALFIT sample and 14 to 88~pc for the CI-estimated sample. In Figure~\ref{jacobi-ratio}, we plot the distribution of half-mass to Jacobi radii ratios for the clusters in NGC~628. The blue histogram shows clusters in the GALFIT sample and the orange histogram shows those in the CI-estimated sample. We find that the ratio $r_{\mathrm{hm}}/r_{\mathrm{J}}$ takes on values between $\sim$0.02 and 0.6 (0.7 for the CI-estimated sample). About 40\% of the clusters have $r_{\mathrm{hm}}/r_{\mathrm{J}}<0.15$, and 60\% have $r_{\mathrm{hm}}/r_{\mathrm{J}}<0.2$, meaning that about half of the clusters are tidally-underfilling while the other half are filling their Roche volumes. We also find a slight, but significant, according to Kendall's $\tau$, positive correlation between cluster age and $r_{\mathrm{hm}}/r_{\mathrm{J}}$, suggesting that the older clusters in NGC~628 are more likely to be tidally-filling. These results are very similar to those from M83 in Paper~I.

\section{Discussion and Conclusions}
\label{discussion-conclusions}

In this work, we have built upon the findings for YMCs in M83 described in Paper~I using YMC samples from two galaxies in the LEGUS survey, NGC~628 and NGC~1313. Using similar techniques and datasets, we find strong agreement between the distributions of effective radii and EFF power-law index for YMCs in these three galaxies. We also find very similar results when comparing the structural parameters to other cluster properties, such as age, mass, and galactocentric distance. In addition, we introduce a new technique to measure effective radii of YMCs which utilizes the concentration index (CI) and agrees well with the GALFIT method for fitting radial intensity profiles. Although the YMCs in M83 span a slightly different age and mass range than those in NGC~628 and NGC~1313, the similarity of our results for these three systems suggest that the galaxy environment has little effect on the structural parameters of the YMC populations.

One important result of this work is that the vast majority of YMCs with ages of $\geq$10~Myr for which we are able to measure effective radii are very likely to be gravitationally bound at the present time, given that their ages exceed their dynamical times by a significant margin (Section~\ref{dynamical-age}). This result shows that the sample selection techniques used by LEGUS, and especially the visual inspection step, which rejects many spurious sources, are successful at producing catalogs of bona fide, bound star clusters (visual inspection class 1 and 2). Much of the disagreement in the literature over the properties of cluster populations in nearby galaxies has stemmed from the methods used to define the samples for study. Our results strongly support the notion that careful selection of centrally-concentrated, symmetric sources that are extended relative to the stellar PSF is the best method for finding bound, potentially long-lived star clusters.

Because the structures of YMCs derived from this work are similar to those for M83, the astrophysical implications are similar as well. First, it is likely that many of the YMCs in the two LEGUS galaxies are not tidally-limited. We find that estimates of the Jacobi radii suggest that about half of the clusters in NGC~628 are underfilling their Roche volumes, while the other half appear to be filling them, and that they appear to be more likely to fill their Roche volumes as they get older. Therefore, the tidal field of NGC~628 may be influencing the structure of some of its YMCs, but overall, the effect appears to be minor. In NGC~1313, we are unable to estimate Jacobi radii, but the other results indicate the tidal field also does not have a significant influence on its clusters.

The effective radii of clusters that are not tidally-limited can expand gradually under the influence of stellar mass loss, two-body relaxation, and possibly interactions with giant molecular clouds \citep{heggiehut2003, portegieszwart2010}. Consistent with this expectation, we observe an overall increase in median radius as our clusters age from a few to 200~Myr old, suggesting that many of the star clusters in our samples are still expanding to fill their tidal radii. Similarly, we also would expect to see a stronger correlation between effective radius and galactocentric distance if  star clusters were significantly affected by the tidal field of their host galaxy \citep[e.g.,][]{madrid2012,puzia2014}. 

Recently, the combination of tidal shocks by passing giant molecular clouds (GMCs) and two-body relaxation have been invoked to explain the near constancy of effective radii in YMCs by \cite{gieles2016}. Two-body relaxation causes clusters to expand while GMC interactions act to make them contract in a balancing act that results in a near-constant effective radius as a function of cluster mass, $r_{\mathrm{hm}}\propto M^{1/9}$. Including other effects, such as stellar mass loss and binary stars, would be an interesting next step. We find weak relationships between radius and mass for clusters in both of the LEGUS galaxies studied here, in general agreement with the prediction of \cite{gieles2016}. Assuming YMCs are modern-day proto-globular clusters, the striking similarity between their distributions of effective radii implies that some mechanisms must balance to allow bound star clusters to remain roughly the same size for very long timescales.

\acknowledgments
These observations are associated with program \# 13364.
Support for program \# 13364 was provided by NASA through a
grant from the Space Telescope Science Institute.
D.A.G. kindly acknowledges financial support by the German Research Foundation (DFG) through grant GO1659/3-2. M.F. acknowledges support by the Science and Technology Facilities Council (grant number ST/L00075X/1).

\software{SourceExtractor \citep{bertin1996}, Yggdrasil \citep{zackrisson2011}, Starburst99 \citep{leitherer1999}, GALFIT \citep{peng2002, peng2010}, IRAF, scipy, baolab \citep{larsen1999b}}

%{\it Facilities:} 
\facility{HST (ACS/WFC, WFC3/UVIS)}

\bibliographystyle{aasjournal}
\bibliography{legus_sizes.bib}

\newcommand{\noop}[1]{}
\begin{thebibliography}{}
\expandafter\ifx\csname natexlab\endcsname\relax\def\natexlab#1{#1}\fi
\providecommand{\url}[1]{\href{#1}{#1}}

\bibitem[{{Adamo} {et~al.}(2017){Adamo}, {Ryon}, {Messa}, {Kim}, {Grasha},
  {Cook}, {Calzetti}, {Lee}, {Whitmore}, {Elmegreen}, {Ubeda}, {Smith},
  {Bright}, {Runnholm}, {Andrews}, {Fumagalli}, {Gouliermis}, {Kahre}, {Nair},
  {Thilker}, {Walterbos}, {Wofford}, {Aloisi}, {Ashworth}, {Brown}, {Chandar},
  {Christian}, {Cignoni}, {Clayton}, {Dale}, {de Mink}, {Dobbs}, {Elmegreen},
  {Evans}, {Gallagher}, {Grebel}, {Herrero}, {Hunter}, {Johnson}, {Kennicutt},
  {Krumholz}, {Lennon}, {Levay}, {Martin}, {Nota}, {Ostlin}, {Pellerin},
  {Prieto}, {Regan}, {Sabbi}, {Sacchi}, {Schaerer}, {Shabani}, {Tosi}, {Van
  Dyk}, \& {Zackrisson}}]{adamo2017}
{Adamo}, A., {Ryon}, J.~E., {Messa}, M., {et~al.} 2017, ArXiv e-prints,
  arXiv:1705.01588

\bibitem[{{Alexander} \& {Gieles}(2013)}]{alexander2013}
{Alexander}, P.~E.~R., \& {Gieles}, M. 2013, \mnras, 432, L1

\bibitem[{{Alexander} {et~al.}(2014){Alexander}, {Gieles}, {Lamers}, \&
  {Baumgardt}}]{alexander2014}
{Alexander}, P.~E.~R., {Gieles}, M., {Lamers}, H.~J.~G.~L.~M., \& {Baumgardt},
  H. 2014, \mnras, 442, 1265

\bibitem[{{Barmby} {et~al.}(2006){Barmby}, {Kuntz}, {Huchra}, \&
  {Brodie}}]{barmby2006}
{Barmby}, P., {Kuntz}, K.~D., {Huchra}, J.~P., \& {Brodie}, J.~P. 2006, \aj,
  132, 883

\bibitem[{{Barmby} {et~al.}(2009){Barmby}, {Perina}, {Bellazzini}, {Cohen},
  {Hodge}, {Huchra}, {Kissler-Patig}, {Puzia}, \& {Strader}}]{barmby2009}
{Barmby}, P., {Perina}, S., {Bellazzini}, M., {et~al.} 2009, \aj, 138, 1667

\bibitem[{{Bastian} {et~al.}(2013){Bastian}, {Cabrera-Ziri}, {Davies}, \&
  {Larsen}}]{bastian2013}
{Bastian}, N., {Cabrera-Ziri}, I., {Davies}, B., \& {Larsen}, S.~S. 2013,
  \mnras, 436, 2852

\bibitem[{{Bastian} {et~al.}(2005){Bastian}, {Gieles}, {Lamers}, {Scheepmaker},
  \& {de Grijs}}]{bastian2005}
{Bastian}, N., {Gieles}, M., {Lamers}, H.~J.~G.~L.~M., {Scheepmaker}, R.~A., \&
  {de Grijs}, R. 2005, \aap, 431, 905

\bibitem[{{Bastian} {et~al.}(2012){Bastian}, {Adamo}, {Gieles}, {Silva-Villa},
  {Lamers}, {Larsen}, {Smith}, {Konstantopoulos}, \&
  {Zackrisson}}]{bastian2012a}
{Bastian}, N., {Adamo}, A., {Gieles}, M., {et~al.} 2012, \mnras, 419, 2606

\bibitem[{{Bertin} \& {Arnouts}(1996)}]{bertin1996}
{Bertin}, E., \& {Arnouts}, S. 1996, \aaps, 117, 393

\bibitem[{{Calzetti} {et~al.}(2015){Calzetti}, {Lee}, {Sabbi}, {Adamo},
  {Smith}, {Andrews}, {Ubeda}, {Bright}, {Thilker}, {Aloisi}, {Brown},
  {Chandar}, {Christian}, {Cignoni}, {Clayton}, {da Silva}, {de Mink}, {Dobbs},
  {Elmegreen}, {Elmegreen}, {Evans}, {Fumagalli}, {Gallagher}, {Gouliermis},
  {Grebel}, {Herrero}, {Hunter}, {Johnson}, {Kennicutt}, {Kim}, {Krumholz},
  {Lennon}, {Levay}, {Martin}, {Nair}, {Nota}, {{\"O}stlin}, {Pellerin},
  {Prieto}, {Regan}, {Ryon}, {Schaerer}, {Schiminovich}, {Tosi}, {Van Dyk},
  {Walterbos}, {Whitmore}, \& {Wofford}}]{calzetti2015}
{Calzetti}, D., {Lee}, J.~C., {Sabbi}, E., {et~al.} 2015, \aj, 149, 51

\bibitem[{{Cardelli} {et~al.}(1989){Cardelli}, {Clayton}, \&
  {Mathis}}]{cardelli1989}
{Cardelli}, J.~A., {Clayton}, G.~C., \& {Mathis}, J.~S. 1989, \apj, 345, 245

\bibitem[{{Daigle} {et~al.}(2006){Daigle}, {Carignan}, {Amram}, {Hernandez},
  {Chemin}, {Balkowski}, \& {Kennicutt}}]{daigle2006}
{Daigle}, O., {Carignan}, C., {Amram}, P., {et~al.} 2006, \mnras, 367, 469

\bibitem[{{de Vaucouleurs}(1963)}]{devaucouleurs1963}
{de Vaucouleurs}, G. 1963, \apj, 137, 720

\bibitem[{{Elson} {et~al.}(1987){Elson}, {Fall}, \& {Freeman}}]{elson1987}
{Elson}, R.~A.~W., {Fall}, S.~M., \& {Freeman}, K.~C. 1987, \apj, 323, 54

\bibitem[{{Fall} \& {Chandar}(2012)}]{fall2012}
{Fall}, S.~M., \& {Chandar}, R. 2012, \apj, 752, 96

\bibitem[{{Feigelson} \& {Jogesh Babu}(2012)}]{feigelson2012}
{Feigelson}, E.~D., \& {Jogesh Babu}, G. 2012, {Modern Statistical Methods for
  Astronomy}

\bibitem[{{Fouesneau} \& {Lan{\c c}on}(2010)}]{fouesneau2010}
{Fouesneau}, M., \& {Lan{\c c}on}, A. 2010, \aap, 521, A22

\bibitem[{{Gieles} {et~al.}(2011){Gieles}, {Heggie}, \& {Zhao}}]{gieles2011b}
{Gieles}, M., {Heggie}, D.~C., \& {Zhao}, H. 2011, \mnras, 413, 2509

\bibitem[{{Gieles} \& {Portegies Zwart}(2011)}]{gieles2011a}
{Gieles}, M., \& {Portegies Zwart}, S.~F. 2011, \mnras, 410, L6

\bibitem[{{Gieles} {et~al.}(2006){Gieles}, {Portegies Zwart}, {Baumgardt},
  {Athanassoula}, {Lamers}, {Sipior}, \& {Leenaarts}}]{gieles2006c}
{Gieles}, M., {Portegies Zwart}, S.~F., {Baumgardt}, H., {et~al.} 2006, \mnras,
  371, 793

\bibitem[{{Gieles} \& {Renaud}(2016)}]{gieles2016}
{Gieles}, M., \& {Renaud}, F. 2016, ArXiv e-prints, arXiv:1605.05940

\bibitem[{{Glatt} {et~al.}(2009){Glatt}, {Grebel}, {Gallagher}, {Nota},
  {Sabbi}, {Sirianni}, {Clementini}, {Da Costa}, {Tosi}, {Harbeck}, {Koch}, \&
  {Kayser}}]{glatt2009}
{Glatt}, K., {Grebel}, E.~K., {Gallagher}, III, J.~S., {et~al.} 2009, \aj, 138,
  1403

\bibitem[{{Harris}(2009)}]{harris2009}
{Harris}, W.~E. 2009, \apj, 699, 254

\bibitem[{{Harris} {et~al.}(2010){Harris}, {Spitler}, {Forbes}, \&
  {Bailin}}]{harris2010}
{Harris}, W.~E., {Spitler}, L.~R., {Forbes}, D.~A., \& {Bailin}, J. 2010,
  \mnras, 401, 1965

\bibitem[{{Heggie} \& {Hut}(2003)}]{heggiehut2003}
{Heggie}, D., \& {Hut}, P. 2003, {The Gravitational Million-Body Problem: A
  Multidisciplinary Approach to Star Cluster Dynamics}

\bibitem[{{H{\'e}nault-Brunet} {et~al.}(2012){H{\'e}nault-Brunet}, {Gieles},
  {Evans}, {Sana}, {Bastian}, {Ma{\'{\i}}z Apell{\'a}niz}, {Taylor}, {Markova},
  {Bressert}, {de Koter}, \& {van Loon}}]{henaultbrunet2012}
{H{\'e}nault-Brunet}, V., {Gieles}, M., {Evans}, C.~J., {et~al.} 2012, \aap,
  545, L1

\bibitem[{{H{\'e}non}(1961)}]{henon1961}
{H{\'e}non}, M. 1961, Annales d'Astrophysique, 24, 369

\bibitem[{{Holtzman} {et~al.}(1996){Holtzman}, {Watson}, {Mould}, {Gallagher},
  {Ballester}, {Burrows}, {Clarke}, {Crisp}, {Evans}, {Griffiths}, {Hester},
  {Hoessel}, {Scowen}, {Stapelfeldt}, {Trauger}, \& {Westphal}}]{holtzman1996}
{Holtzman}, J.~A., {Watson}, A.~M., {Mould}, J.~R., {et~al.} 1996, \aj, 112,
  416

\bibitem[{{Ivezi{\'c}} {et~al.}(2014){Ivezi{\'c}}, {Connelly}, {VanderPlas}, \&
  {Gray}}]{ivezic2014}
{Ivezi{\'c}}, {\v Z}., {Connelly}, A.~J., {VanderPlas}, J.~T., \& {Gray}, A.
  2014, {Statistics, Data Mining, and Machine Learning in Astronomy}

\bibitem[{{Jacobs} {et~al.}(2009){Jacobs}, {Rizzi}, {Tully}, {Shaya},
  {Makarov}, \& {Makarova}}]{jacobs2009}
{Jacobs}, B.~A., {Rizzi}, L., {Tully}, R.~B., {et~al.} 2009, \aj, 138, 332

\bibitem[{{Jord{\'a}n} {et~al.}(2005){Jord{\'a}n}, {C{\^o}t{\'e}}, {Blakeslee},
  {Ferrarese}, {McLaughlin}, {Mei}, {Peng}, {Tonry}, {Merritt},
  {Milosavljevi{\'c}}, {Sarazin}, {Sivakoff}, \& {West}}]{jordan2005}
{Jord{\'a}n}, A., {C{\^o}t{\'e}}, P., {Blakeslee}, J.~P., {et~al.} 2005, \apj,
  634, 1002

\bibitem[{{Kissler-Patig} {et~al.}(2006){Kissler-Patig}, {Jord{\'a}n}, \&
  {Bastian}}]{kisslerpatig2006}
{Kissler-Patig}, M., {Jord{\'a}n}, A., \& {Bastian}, N. 2006, \aap, 448, 1031

\bibitem[{{Larsen}(1999)}]{larsen1999b}
{Larsen}, S.~S. 1999, \aaps, 139, 393

\bibitem[{{Larsen}(2004)}]{larsen2004}
---. 2004, \aap, 416, 537

\bibitem[{{Leitherer} {et~al.}(1999){Leitherer}, {Schaerer}, {Goldader},
  {Delgado}, {Robert}, {Kune}, {de Mello}, {Devost}, \&
  {Heckman}}]{leitherer1999}
{Leitherer}, C., {Schaerer}, D., {Goldader}, J.~D., {et~al.} 1999, \apjs, 123,
  3

\bibitem[{{Mackey} \& {Gilmore}(2003)}]{mackey2003a}
{Mackey}, A.~D., \& {Gilmore}, G.~F. 2003, \mnras, 338, 85

\bibitem[{{Madrid} {et~al.}(2012){Madrid}, {Hurley}, \& {Sippel}}]{madrid2012}
{Madrid}, J.~P., {Hurley}, J.~R., \& {Sippel}, A.~C. 2012, \apj, 756, 167

\bibitem[{{Massey} {et~al.}(2006){Massey}, {Olsen}, {Hodge}, {Strong},
  {Jacoby}, {Schlingman}, \& {Smith}}]{massey2006}
{Massey}, P., {Olsen}, K.~A.~G., {Hodge}, P.~W., {et~al.} 2006, \aj, 131, 2478

\bibitem[{{Masters} {et~al.}(2010){Masters}, {Jord{\'a}n}, {C{\^o}t{\'e}},
  {Ferrarese}, {Blakeslee}, {Infante}, {Peng}, {Mei}, \& {West}}]{masters2010}
{Masters}, K.~L., {Jord{\'a}n}, A., {C{\^o}t{\'e}}, P., {et~al.} 2010, \apj,
  715, 1419

\bibitem[{{Matthews} {et~al.}(1999){Matthews}, {Gallagher}, {Krist}, {Watson},
  {Burrows}, {Griffiths}, {Hester}, {Trauger}, {Ballester}, {Clarke}, {Crisp},
  {Evans}, {Hoessel}, {Holtzman}, {Mould}, {Scowen}, {Stapelfeldt}, \&
  {Westphal}}]{Matthews99}
{Matthews}, L.~D., {Gallagher}, III, J.~S., {Krist}, J.~E., {et~al.} 1999, \aj,
  118, 208

\bibitem[{{McLaughlin} \& {van der Marel}(2005)}]{mclaughlin2005}
{McLaughlin}, D.~E., \& {van der Marel}, R.~P. 2005, \apjs, 161, 304

\bibitem[{{Mora} {et~al.}(2009){Mora}, {Larsen}, {Kissler-Patig}, {Brodie}, \&
  {Richtler}}]{mora2009}
{Mora}, M.~D., {Larsen}, S.~S., {Kissler-Patig}, M., {Brodie}, J.~P., \&
  {Richtler}, T. 2009, \aap, 501, 949

\bibitem[{{Olivares} {et~al.}(2010){Olivares}, {Hamuy}, {Pignata}, {Maza},
  {Bersten}, {Phillips}, {Suntzeff}, {Filippenko}, {Morrel}, {Kirshner}, \&
  {Matheson}}]{olivares2010}
{Olivares}, E., F., {Hamuy}, M., {Pignata}, G., {et~al.} 2010, \apj, 715, 833

\bibitem[{{Pellerin} {et~al.}(2007){Pellerin}, {Meyer}, {Harris}, \&
  {Calzetti}}]{pellerin2007}
{Pellerin}, A., {Meyer}, M., {Harris}, J., \& {Calzetti}, D. 2007, \apjl, 658,
  L87

\bibitem[{{Peng} {et~al.}(2002){Peng}, {Ho}, {Impey}, \& {Rix}}]{peng2002}
{Peng}, C.~Y., {Ho}, L.~C., {Impey}, C.~D., \& {Rix}, H.-W. 2002, \aj, 124, 266

\bibitem[{{Peng} {et~al.}(2010){Peng}, {Ho}, {Impey}, \& {Rix}}]{peng2010}
---. 2010, \aj, 139, 2097

\bibitem[{{Peters} {et~al.}(1994){Peters}, {Freeman}, {Forster}, {Manchester},
  \& {Ables}}]{peters1994}
{Peters}, W.~L., {Freeman}, K.~C., {Forster}, J.~R., {Manchester}, R.~N., \&
  {Ables}, J.~G. 1994, \mnras, 269, 1025

\bibitem[{{Popescu} \& {Hanson}(2010)}]{popescu2010}
{Popescu}, B., \& {Hanson}, M.~M. 2010, \apj, 724, 296

\bibitem[{{Portegies Zwart} {et~al.}(2010){Portegies Zwart}, {McMillan}, \&
  {Gieles}}]{portegieszwart2010}
{Portegies Zwart}, S.~F., {McMillan}, S.~L.~W., \& {Gieles}, M. 2010, \araa,
  48, 431

\bibitem[{{Puzia} {et~al.}(2014){Puzia}, {Paolillo}, {Goudfrooij}, {Maccarone},
  {Fabbiano}, \& {Angelini}}]{puzia2014}
{Puzia}, T.~H., {Paolillo}, M., {Goudfrooij}, P., {et~al.} 2014, \apj, 786, 78

\bibitem[{{Ryder} {et~al.}(1995){Ryder}, {Staveley-Smith}, {Malin}, \&
  {Walsh}}]{ryder1995}
{Ryder}, S.~D., {Staveley-Smith}, L., {Malin}, D., \& {Walsh}, W. 1995, \aj,
  109, 1592

\bibitem[{{Ryon} {et~al.}(2015){Ryon}, {Bastian}, {Adamo}, {Konstantopoulos},
  {Gallagher}, {Larsen}, {Hollyhead}, {Silva-Villa}, \& {Smith}}]{ryon2015}
{Ryon}, J.~E., {Bastian}, N., {Adamo}, A., {et~al.} 2015, \mnras, 452, 525

\bibitem[{{Scheepmaker} {et~al.}(2007){Scheepmaker}, {Haas}, {Gieles},
  {Bastian}, {Larsen}, \& {Lamers}}]{scheepmaker2007}
{Scheepmaker}, R.~A., {Haas}, M.~R., {Gieles}, M., {et~al.} 2007, \aap, 469,
  925

\bibitem[{{Silva-Villa} \& {Larsen}(2011)}]{silvavilla2011}
{Silva-Villa}, E., \& {Larsen}, S.~S. 2011, \aap, 529, A25

\bibitem[{{Silva-Villa} \& {Larsen}(2012)}]{silvavilla2012}
---. 2012, \mnras, 423, 213

\bibitem[{{Smith} \& {Gallagher}(2001)}]{Smith01}
{Smith}, L.~J., \& {Gallagher}, J.~S. 2001, \mnras, 326, 1027

\bibitem[{{Sun} {et~al.}(2016){Sun}, {de Grijs}, {Fan}, \& {Cameron}}]{Sun16}
{Sun}, W., {de Grijs}, R., {Fan}, Z., \& {Cameron}, E. 2016, \apj, 816, 9

\bibitem[{{Thilker} {et~al.}(2007){Thilker}, {Bianchi}, {Meurer}, {Gil de Paz},
  {Boissier}, {Madore}, {Boselli}, {Ferguson}, {Mu{\~n}oz-Mateos}, {Madsen},
  {Hameed}, {Overzier}, {Forster}, {Friedman}, {Martin}, {Morrissey}, {Neff},
  {Schiminovich}, {Seibert}, {Small}, {Wyder}, {Donas}, {Heckman}, {Lee},
  {Milliard}, {Rich}, {Szalay}, {Welsh}, \& {Yi}}]{thilker2007}
{Thilker}, D.~A., {Bianchi}, L., {Meurer}, G., {et~al.} 2007, \apjs, 173, 538

\bibitem[{{Trenti} {et~al.}(2010){Trenti}, {Vesperini}, \&
  {Pasquato}}]{Trenti10}
{Trenti}, M., {Vesperini}, E., \& {Pasquato}, M. 2010, \apj, 708, 1598

\bibitem[{{van den Bergh} {et~al.}(1991){van den Bergh}, {Morbey}, \&
  {Pazder}}]{vandenbergh1991}
{van den Bergh}, S., {Morbey}, C., \& {Pazder}, J. 1991, \apj, 375, 594

\bibitem[{{V{\'a}zquez} \& {Leitherer}(2005)}]{vazquez2005}
{V{\'a}zquez}, G.~A., \& {Leitherer}, C. 2005, \apj, 621, 695

\bibitem[{{Webb} {et~al.}(2016){Webb}, {Sills}, {Harris}, {G{\'o}mez},
  {Paolillo}, {Woodley}, \& {Puzia}}]{webb2016}
{Webb}, J.~J., {Sills}, A., {Harris}, W.~E., {et~al.} 2016, ArXiv e-prints,
  arXiv:1605.03191

\bibitem[{{Whitmore} {et~al.}(1999){Whitmore}, {Zhang}, {Leitherer}, {Fall},
  {Schweizer}, \& {Miller}}]{whitmore1999}
{Whitmore}, B.~C., {Zhang}, Q., {Leitherer}, C., {et~al.} 1999, \aj, 118, 1551

\bibitem[{{Whitmore} {et~al.}(2010){Whitmore}, {Chandar}, {Schweizer},
  {Rothberg}, {Leitherer}, {Rieke}, {Rieke}, {Blair}, {Mengel}, \&
  {Alonso-Herrero}}]{whitmore2010}
{Whitmore}, B.~C., {Chandar}, R., {Schweizer}, F., {et~al.} 2010, \aj, 140, 75

\bibitem[{{Zackrisson} {et~al.}(2011){Zackrisson}, {Rydberg}, {Schaerer},
  {{\"O}stlin}, \& {Tuli}}]{zackrisson2011}
{Zackrisson}, E., {Rydberg}, C.-E., {Schaerer}, D., {{\"O}stlin}, G., \&
  {Tuli}, M. 2011, \apj, 740, 13

\bibitem[{{Zepf} {et~al.}(1999){Zepf}, {Ashman}, {English}, {Freeman}, \&
  {Sharples}}]{zepf1999}
{Zepf}, S.~E., {Ashman}, K.~M., {English}, J., {Freeman}, K.~C., \& {Sharples},
  R.~M. 1999, \aj, 118, 752

\end{thebibliography}

\appendix

\section{Kendall's $\tau$}
\label{appendix}
We use the Kendall's $\tau$ correlation test extensively in this paper, and so a more in-depth explanation of its definition and use is warranted. The following draws heavily from \cite{feigelson2012} and \cite{ivezic2014}. The purpose of the Kendall's $\tau$ correlation test is to determine if a correlation exists between sets of paired measurements, $(X_1,Y_1),...,(X_n,Y_n)$. This is achieved by comparing the number of concordant pairs, for which $(Y_i - Y_j)/(X_i - X_j)>0$, with the number of discordant pairs, for which $(Y_i - Y_j)/(X_i - X_j) < 0$. Essentially, to be counted as a concordant pair, the differences in the $X$ and $Y$ values must have the same sign. To be counted as a discordant pair, the differences in the $X$ and $Y$ values must have different signs. If the paired measurements were perfectly correlated (anticorrelated), all of the pairs would be concordant (discordant). If the $X$ values or the $Y$ values are equal for a given pair, this is called a tie.

We use the \texttt{kendalltau} function within the \texttt{stats} module of the python package \texttt{scipy} to perform the correlation tests. This function defines Kendall's $\tau$ as
\begin{equation}
\tau = \frac{P-Q}{\sqrt{(P+Q+T)(P+Q+U)}},
\end{equation}
where $P$ is the number of concordant pairs, $Q$ is the number of discordant pairs, $T$ is the number of ties only in $X$, and $U$ is the number of ties only in $Y$. If a tie occurs in both the $X$ values and the $Y$ values for the same pair, it is not added to either $T$ or $U$. If all of the pairs are concordant and there are no ties, then $\tau=1$, implying perfect correlation. If all of the pairs are discordant and there are no ties, then $\tau = -1$, implying perfect anticorrelation. As stated by \cite{ivezic2014}, Kendall's $\tau$ can be interpreted as the probability that the two datasets, $X$ and $Y$, are in the same order minus the probability that they are not in the same order.

We choose the Kendall's $\tau$ correlation test because it is nonparametric, i.e. it does not require knowledge of the underlying distributions of the data, nor does it assume that any correlation between the two datasets will be linear. Pearson's correlation coefficient, $r$, assumes the data are sampled from a bivariate Gaussian distribution, and only looks for linear correlations. Spearman's $\rho$ correlation test is similar in nature to Kendall's $\tau$, and gives very similar results, but does not approach normality as quickly for small samples as Kendall's $\tau$. 

An example of the use of Kendall's $\tau$ correlation test is given in Section~8.8.1 in \cite{feigelson2012}. They present a dataset containing 20 properties of 147 Milky Way globular clusters, and look for a correlation between each pair of properties with Kendall's $\tau$. They find a variety of correlations, both positive and negative, and many are nonlinear. Over half of the correlations are statistically significant, with $\left|\tau\right|>0.2$ and $p < 0.003$, corresponding to $>3\sigma$ relationships.

\end{document}